\documentclass[prd,amsmath,amssymb,preprintnumbers,superscriptaddress,twocolumn,10pt,nofootinbib,showkeys,showpacs]{revtex4}

\pdfoutput=1

\usepackage{graphicx}
\usepackage{dcolumn}
\usepackage{bm}
\usepackage{amssymb}
\usepackage{latexsym}
\usepackage{booktabs}
\usepackage{amsmath}
\usepackage{multirow}
\usepackage{url}

\usepackage{float}
\usepackage[colorlinks=true, linkcolor=red, citecolor=blue]{hyperref}

\usepackage[normalem]{ulem}
\usepackage{color}
\usepackage{array}
\usepackage{enumerate}

\begin{document}

\newcommand{\dm}{{\rm DM}}
\newcommand{\DM}{{\rm DM}}
\newcommand{\pccc}{\,pc\,cm$^{-3}$\,}

\title{Combining strongly lensed and unlensed fast radio bursts: To be a more precise late-universe probe}

\author{Ji-Guo Zhang}
\affiliation{Key Laboratory of Cosmology and Astrophysics (Liaoning Province) \& Department of Physics, College of Sciences, Northeastern University, Shenyang 110819, China}
\author{Yi-Fan Jiang}
\affiliation{Key Laboratory of Cosmology and Astrophysics (Liaoning Province) \& Department of Physics, College of Sciences, Northeastern University, Shenyang 110819, China}
\author{Ze-Wei Zhao}
\affiliation{Key Laboratory of Cosmology and Astrophysics (Liaoning Province) \& Department of Physics, College of Sciences, Northeastern University, Shenyang 110819, China}
\author{Jing-Zhao Qi}
\affiliation{Key Laboratory of Cosmology and Astrophysics (Liaoning Province) \& Department of Physics, College of Sciences, Northeastern University, Shenyang 110819, China}
\author{Jing-Fei Zhang}
\affiliation{Key Laboratory of Cosmology and Astrophysics (Liaoning Province) \& Department of Physics, College of Sciences, Northeastern University, Shenyang 110819, China}
\author{Xin Zhang}\thanks{Corresponding author.\\zhangxin@mail.neu.edu.cn}
\affiliation{Key Laboratory of Cosmology and Astrophysics (Liaoning Province) \& Department of Physics, College of Sciences, Northeastern University, Shenyang 110819, China}
\affiliation{Key Laboratory of Data Analytics and Optimization for Smart Industry (Ministry of Education), Northeastern University, Shenyang 110819, China}
\affiliation{National Frontiers Science Center for Industrial Intelligence and Systems Optimization, Northeastern University, Shenyang 110819, China}

\begin{abstract}
The Macquart relation and time-delay cosmography are now two promising ways to fast radio burst (FRB) cosmology. In this work, we propose a joint method that combines strongly lensed and unlensed FRBs for improving cosmological parameter estimation by using simulated FRB data from the future sensitive coherent all-sky monitor survey, which is expected to detect a large number of FRBs including galaxy-galaxy strongly lensed events. We find that using a detectable sample of 100,000 localized FRBs including $40$ lensed events can simultaneously constrain the Hubble constant and the equation of state of dark energy, with high precision of $\varepsilon(H_0)=0.4\%$ and $\varepsilon(w)=4.5\%$ in the simplest dynamical dark energy model. The joint analysis of unlensed and lensed FRBs significantly improves the constraint on $H_0$, which could be more effective than combining either the unlensed FRBs with future gravitational wave (GW) standard sirens or the lensed FRBs with CMB. Furthermore, combining the full FRB sample with the CMB+BAO+SNe data yields $\sigma(H_0)=0.29~{\rm km~s^{-1}~Mpc^{-1}}$, $\sigma(w_0)=0.046$, and $\sigma(w_a)=0.15$ in the two-parameter dynamical dark energy model, which outperform the results from the CMB+BAO+SNe+GW data. This reinforces the cosmological implications of a multi-wavelength observational strategy in optical and radio bands. We conclude that the future FRB observations will shed light on the nature of dark energy and also the Hubble tension if enough events with long-duration lensing are incorporated.
\end{abstract}


\maketitle
\section{Introduction}\label{sec:3}
Cosmology now stands in the midst of a golden age. It is mainly attributed to the exquisite precision in measuring the power spectrum of temperature anisotropies of the cosmic microwave background (CMB), which has ushered in the era of precision cosmology \cite{WMAP:2003elm,WMAP:2003ivt}.
As the standard model of cosmology, the $\Lambda$ cold dark matter ($\Lambda$CDM) model has only six basic parameters but accurately fits most observations, particularly the CMB anisotropies \cite{2020Planck}. 
However, within the standard cosmological scenario, there are still many unsettled issues like the nature of dark energy and some cosmological tensions.

Dark energy is a component with negative pressure that drives the accelerated expansion of the late universe, and understanding its essence requires determining its equation of state (EoS).
The standard $\Lambda$CDM model describes dark energy as a cosmological constant $\Lambda$ with the EoS $w = -1$, which actually suffers from several theoretical problems \cite{Weinberg:1988cp}. Therefore, one has widely proposed dynamical dark energy with the EoS deviating from $w = -1$ or evolving over time \cite{Joyce:2014kja}.
To accurately measure this EoS, low-redshift measurements are employed since the CMB is an early-universe probe, which cannot effectively constrain the extra parameters describing the EoS of dynamical dark energy.
High precision is desirable since the most stringent constraint today is still far away from deciphering dark energy, with the combination of three mainstream observations (CMB+BAO+SNe, where BAO and SNe refer to the observations of baryon acoustic oscillation and type Ia supernovae, respectively) \cite{DESI:2024mwx}. 
Worse still, the tension between the values of the Hubble constant $H_0$ estimated by the early- \cite{2020Planck} and late-universe observations \cite{Riess:2021jrx} has now exceeded $5\sigma$, widely discussed as the ``Hubble tension'' \cite{Riess:2021jrx,Verde:2019ivm,Hu:2023jqc}. No reliable evidence of systematic errors has been found \cite{Follin:2017ljs}, and no extended cosmological model can truly resolve the crisis \cite{Yang:2018euj,Guo:2018ans,Zhang:2019cww,Feng:2019jqa,Liu:2019awo,vag2020new,Gao:2021xnk,Cai:2021wgv,Vagnozzi:2023nrq}. So developing late-universe precise probes is essential for addressing the cosmological issues of both the Hubble tension and dark energy \cite{Moresco:2022phi,Wu:2022dgy}. 
In the coming decades, some novel late-universe probes will be vigorously developed via gravitational-wave (GW) \cite{Zhang:2019ylr} and radio astronomy \cite{Xu:2020uws,Wu:2022jkf}.
In this work, we wish to address the issues by employing the future fast radio burst (FRB) observations for simultaneously measuring the Hubble constant and dynamical dark energy.

FRBs --- bright, millisecond-duration radio pulses at cosmological distances \cite{lorimer2007} --- are the latest large puzzle in the universe and have been attracting intense observational and theoretical investigations in recent years \cite{Bailes:2022pxa,Zhang:2022uzl}.
The FRB sample size is rapidly increasing, primarily due to the contributions of the Canadian Hydrogen Intensity Mapping Experiment Fast Radio Burst project (CHIME/FRB) \cite{bandura2014canadian} and the Five hundred meter Aperture Spherical radio Telescope (FAST) telescope, which have detected the most FRB sources \cite{CHIMEFRB:2021srp} and bursts \cite{Li:2021hpl,Xu:2021qdn,wuziwei:2023}, respectively.
In spite of unclear physical origins, FRBs with known redshifts measured from precisely localized host galaxies have been widely proposed as a cosmological probe \cite{dai2023ska}, owing to the high event rate and detections of increasing localized FRBs (see Refs.~\cite{Bhandari:2021thi,Xiao:2021omr,Wu:2024iyu} for recent reviews). Probing the universe with FRBs can be primarily achieved by two proposed methods --- the ``Macquart relation'' and gravitational lensing analysis.

One is realized by the Macquart relation \cite{Macquart:2020lln}, which makes a connection between the intergalactic medium (IGM) dispersion measure (DM) and redshift \cite{Zhou:2014yta,Gao:2014iva,James:2021jbo}. Characterized by the integrated number density of free electrons along FRB paths, DM records both cosmic evolution and baryonic information across cosmological distance as {\it standard ping} \cite{Masui:2015ola}.
Thus, localized FRBs (with redshifts inferred from identified host galaxies) can be harnessed to determine cosmological parameters,
including those associated with dark energy and the Hubble constant.
For measuring dark energy effectively via the Macquart relation, it is important to accurately extract $\rm DM_{\rm IGM}$ from the total DM, and thus important to quantify the IGM inhomogeneity and host or source DM contribution \cite{Kumar:2019qhc}. To achieve this, a large number of well-localized FRB sample (with at least $\sim 10^4$ events) is required. In the Square Kilometre Array (SKA) era, a million localized FRBs as an independent probe could precisely measure dark energy \cite{Zhang:2023gye} and explore the epoch of reionization \cite{Hashimoto:2021tyk,Wei:2024tpv}. Alternatively, it is effective to utilize the combination of FRB with external cosmological probes like CMB \cite{Zhao:2020ole,Qiu:2021cww,Zhao:2022bpd}, BAO \cite{Zhou:2014yta}, SNe \cite{Gao:2014iva,Jaro:2018vgh}, GW associations \cite{Wei:2018cgd}, the CMB+BAO+SNe+$H_0$ combination \cite{Walters:2017afr}, and information of large scale structure \cite{Zhu:2022mzv} to break parameter inherent degeneracies, which suggests FRBs a sound probe to complement. 
For measuring the Hubble constant, effective constraints often come from the joint analysis of FRB data and big bang nucleosynthesis (BBN) results. For example, various localized FRB datasets were used to constrain $H_{0}$ \cite{Hagstotz:2021jzu,Wu:2021jyk,James:2022dcx, Wei:2023avr,Fortunato:2024hfm,Kalita:2024xae}; recently, Ref.~\cite{Zhao:2022yiv} also developed a Bayesian method to using unlocalized FRBs. 
In addition, combinations of FRB data with SNe datasets \cite{Liu:2022bmn} and with Hubble parameter $H(z)$ measurements \cite{Gao:2023izj} were also explored.

Another prospect of FRB cosmology is to study the gravitational lensing.
The high rate of FRB events suggests the potential of detecting lensed FRBs in future blind surveys. The events strongly lensed by massive galaxies --- referred to as galaxy-galaxy strongly lensed (GGSL) FRBs --- offer a unique tool to probe cosmology. Due to their short durations, the time delays (TDs) between lensed images can be measured with exceptionally high precision, leading to numerous applications \cite{Li:2017mek,Dai:2017twh,Zitrin:2018let,Liu:2019jka,Wucknitz:2020spz,Adi:2021uuw,Zhao:2021jeb,Er:2022lad,Gao:2022ifq,Xiao:2022hkl,Jiang:2024otl}; in particular, the precise measurement of $H_0$ via a technique known as ``time-delay cosmography'' \cite{Refsdal:1964blz,Birrer:2022chj}. By measuring the angular diameter distances of simulated GGSL FRB sources and lenses, Ref.~\cite{Li:2017mek} demonstrated that using a sample of $10$ lensed repeating FRBs could determine $H_0$ with sub-percent precision.
In the next decade, the GGSL FRB events are expected to be detected through future ultra-widefield FRB surveys, such as coherent all-sky monitors (CASMs). 
With a vast field of view (FoV) and accurate localization capability provided by very long baseline interferometry (VLBI), CASM can perform long-term and high-cadence monitoring, which makes it likely to detect the lensed copy of an FRB signal even after a time delay of several months. With a system-equivalent flux density (SEFD) comparable to CHIME, such a sensitive CASM survey could detect $50,000$--$100,000$ FRBs including $5$--$40$ potential GGSL events during a $5$-year observation \cite{Connor:2022bwl}. Note that we refer to this hypothetical survey as ``CASM'' throughout this paper.

The two methods mentioned above (i.e., the Macquart relation and the time-delay cosmography) are currently the most compelling approaches for using localized FRBs as cosmological probes. By precisely measuring DMs from tremendous FRBs, it is possible to effectively constrain dark-energy EoS parameters. However, this approach has limited effectiveness in constraining $H_{0}$ due to potential parameter degeneracies with the baryon density $\Omega_{\rm b}$ \cite{Walters:2017afr} (thus, previous work introducing the BBN prior is arguably not a purely late-universe result). Conversely, accurate measurement of TDs with GGSL FRBs can provide precise constraints on $H_{0}$, but it cannot independently constrain dark energy evolution, also needing other complementary probes like
CMB and SNe \cite{Liu:2019jka,Zhao:2021jeb}. Therefore, combining these two methodologies, which respectively offer remarkable constraints on dark-energy EoS parameters and $H_0$, has the potential to break mutual degeneracies and merits serious consideration for the realm of FRB cosmology.

In this study, we first combine TD and DM measurements from strongly lensed and unlensed FRBs, respectively, to constrain the late-universe physics. We wish to answer what extent the Hubble constant and the EoS of dark energy can be simultaneously measured using the localized FRB sample (including GGSL events) from the future sensitive CASM survey. We assume that the CASM will build VLBI outriggers to precisely localize the host galaxies of FRBs and determine their redshifts.

This paper is organized as follows. In sect.~\ref{sec2}, we describe the methods of simulating FRB data and other cosmological data used. We show the constraint results in sect.~\ref{sec3} and make a relevant discussion in sect.~\ref{sec4}. The conclusion is given in sect.~\ref{sec5}.

\section{Methods and data}\label{sec2}

\subsection{TD measurement from lensed FRBs} \label{td}

\begin{table}[htbp]
	\renewcommand\arraystretch{1.5}
	\caption{Uncertainties of three components contributing to the uncertainty of TD distance measurements. $\varepsilon(\Delta t)$, $\varepsilon(\Delta \psi)$, and $\varepsilon(\kappa_{\rm ext})$ correspond to the relative uncertainties of time delay, Fermat potential difference, and LOS contamination, respectively.}\label{tab:tab1}
	\centering
	\renewcommand{\arraystretch}{2}
	\setlength{\tabcolsep}{14pt}
	\begin{tabular}{cccc}
		\hline\hline
		GGSL source                    & $\varepsilon(\Delta t)$         & $\varepsilon(\Delta \psi)$            & $\varepsilon (\kappa_{\rm ext})$         \\
		\hline
		Lensed QSOs         & 5\%                  &  3\%                              &  3\%          \\
		Lensed SNe            & 3\%                &  1\%                               &  3\%                \\
		Lensed FRBs         & 0               &  0.8\%                               &  2\%                \\
		\hline\hline
	\end{tabular}
\end{table}

In gravitational lensing, the time delay between the arrival times of photons for images $i$ and $j$ can be predicted as \cite{Birrer:2022chj}
\begin{equation}
	\Delta t_{i,j}=\frac{(1+z_{\rm l})D_{\Delta t}}{c}\Delta \phi_{i,j},
\end{equation}
where $z_{\rm l}$ is the redshift of lens and $c$ is the light speed. The ``time-delay distance'' $D_{\Delta t}$ is defined as \cite{Refsdal:1964blz}
\begin{equation}\label{eq2}
	D_{\Delta t} \equiv \left(1+z_{\rm l}\right) \frac{D_{\rm l}^{\rm A} D_{\rm s}^{\rm A}}{D_{\rm ls}^{\rm A}},
\end{equation}
where $D_{\rm l}^{\rm A}$, $D_{\rm s}^{\rm A}$, and $D_{\rm ls}^{\rm A}$ are the angular diameter distances between observer and lens, between observer and source, and between lens and source, respectively. The Fermat potential difference $\Delta \phi_{i,j}$ is defined as
\begin{equation}
	\Delta \phi_{i,j} = \left[\frac{\left(\boldsymbol{\theta}_{i}-\boldsymbol{\beta}\right)^{2}}{2}-\psi\left(\boldsymbol{\theta}_{i}\right)-\frac{\left(\boldsymbol{\theta}_{j}-\boldsymbol{\beta}\right)^{2}}{2}+\psi\left(\boldsymbol{\theta}_{j}\right)\right],
\end{equation}
where $\boldsymbol{\theta}_{i}$ and $\boldsymbol{\theta}_{j}$ are the angular positions of two images, $\boldsymbol{\beta}$ is the source position, and $\psi$ is the lensing two-dimensional potential related to its mass distribution.

Based on the relationship between the dimensionless comoving distance and the angular diameter distance $d\left(z_{\rm l}, z_{\rm s}\right) \equiv\left(1+z_{\rm s}\right) H_{0} D_{\rm A}\left(z_{\rm l}, z_{\rm s}\right)/c$, we can rewrite Eq.~(\ref{eq2}) as
\begin{equation}\label{eq:deltat}
	D_{\Delta t} =\frac{c}{H_{0}} \frac{d_{\rm l} d_{\rm s}}{d_{\rm ls}}.
\end{equation}
We can see that $D_{\Delta t}$ is inversely proportional to $H_0$. So if we can measure both redshifts and $\Delta \phi_{i,j}$ (of course, including $\boldsymbol{\theta}_{i}$, $\boldsymbol{\theta}_{j}$, and $\boldsymbol{\beta}$) from modeling the observational data, we can measure $H_0$ \cite{Treu:2023mih}. 
This method has been intensively employed to study time-delay cosmography \cite{Wang:2019yob,Wang:2021kxc,Qi:2022sxm,Qi:2022kfg,Li:2023gpp,Li:2023jxc} and fundamental physics \cite{Cao:2018rzc, Qi:2018aio,Liu:2021xvc,Qi:2024acx}.
In this work, we focus on galaxy-scale strongly lensing of FRBs (see Refs.~\cite{Munoz:2016tmg,Laha:2018zav,Liao:2020wae,Zhou:2021ygz,Zhou:2021ndx,Tsai:2023tyw,Xiao:2024qay}, which discuss FRB microlensing scenarios) and assume the singular isothermal sphere (SIS) model for lensing event estimation (see Appendix \ref{Appendix} for more details).

In order to estimate the time-delay distance in Eq.~(\ref{eq2}), we identify three primary sources of uncertainty, i.e., the measurement of the time delay, the reconstruction of the Fermat potential, and modeling the line of sight (LOS) environment.

For a strongly lensed FRB system, the time delay can be measured with ultra-precise precision, since the short duration of the transient ($\sim$ ms) is significantly less than the typical galaxy-lensing time delay ($\sim 10$ d). Thus, the relative uncertainty in TD measurement of strongly lensed FRB sources ($\varepsilon(\Delta t$)) can be considered negligible (i.e., $\varepsilon(\Delta t) = 0$).

The uncertainty related to the Fermat potential (relative uncertainty denoted as $\varepsilon(\Delta \psi)$) depends on lens modeling. The absence of dazzling active galactic nucleus (AGN) contamination within the source galaxy takes advantage for reconstructing the lens mass distribution and obtaining a clear image of the host galaxies in lensed FRB systems.
Through simulations based on HST WFC3 observations from transient sources like FRBs, Ref.~\cite{Ding:2021bxs} found that the precision of the Fermat potential reconstruction could be improved by a factor of $\sim 4$ when comparing lensed transients to lensed AGNs. 
Furthermore, Ref.~\cite{Li:2017mek} showed that lens mass modeling only introduces $\sim 0.8\%$ uncertainty to $D_{\Delta t}$.
However, the precision could be diminished due to the effect of mass-sheet degeneracy (MSD), where different mass models could produce identical strong lensing observables (e.g., image positions) but imply different values of $H_0$ \cite{schneider2013mass}.
It is necessary to note that the $0.8\%$ uncertainty from Ref.~\cite{Li:2017mek} assumes a correctly chosen lens mass profile model, which is challenging to fit accurately in observations and can exhibit degeneracies beyond the MSD \cite{Wagner:2018rae}.
Nonetheless, such degeneracies may be mitigated by leveraging FRBs with their high-precision TD measurements \cite{Wagner:2018fvv}.
In addition, due to the at least sub-arcsecond high localization precision of FRBs with VLBI outrigger stations, they can be treated as point sources in lens modeling, reducing positional uncertainties in determining the foreground lens potential.
In general, based on simulations presented in Ref.~\cite{Li:2017mek}, we adopt a $0.8\%$ relative uncertainty on the Fermat potential for the measurements of $D_{\Delta t}$ (i.e., $\varepsilon( \Delta \psi) = 0.8\%$).

The last component of uncertainty is contributed by LOS environment modeling ($\varepsilon(\kappa_{\rm ext})$). This budget is generally characterized by an external convergence ($\kappa_{\rm ext}$), which is resulted from the excess mass close in projection to the lensing galaxies along the LOS. Taking this effect into account, the actual $D_{\Delta t}$ is corrected to $D_{\Delta t} = {D_{\Delta t}^{\rm model}}/(1 - \kappa_{\rm ext})$. In the case of the lens HE 0435-1223 \cite{Wis:2002dp}, $\varepsilon( \kappa_{\rm ext})$ could be limited to $2.5\%$ through weighted galaxy counts, and a $1.6\%$ uncertainty by utilizing an inpainting technique and multi-scale entropy filtering algorithm \cite{Tih:2017mym}. Therefore, it is reasonable to take a $2.0\%$ relative uncertainty on the LOS environment modeling introduced to $D_{\Delta t}$ for upcoming lensed FRB systems (i.e., $\varepsilon (\kappa_{\rm ext}) = 2.0\%$). 

Overall, the total uncertainty of $D_{\Delta t}$ can be propagated as:
\begin{equation}\label{eq:sigmaDdeltat}
	\sigma_{D_{\Delta t}}=D_{\Delta t} \times \left[\varepsilon^{2}(\Delta t)+{\varepsilon^{2}(\Delta \psi)}
	+{\varepsilon^{2}(\kappa_{\rm ext})}\right]^{1/2}.
\end{equation}
The uncertainty levels of all budgets we adopted are outlined in Table~\ref{tab:tab1}, which also lists the corresponding uncertainties for lensed quasars (QSOs) and lensed SNe for comparison \cite{H0LiCOW:2019xdh,Suyu:2020opl,Qi:2022sxm}.
This shows the advantages of using FRBs for precisely measuring $D_{\Delta t}$.
For a strongly lensed FRB system, $\sigma_{D_{\Delta t}}$ achieves a high precision level of $\sim 2.15\%$ using Eq.~(\ref{eq:sigmaDdeltat}). 
Also, FRBs occur much more frequently than SNe in the universe, so the possibility of FRBs being strongly lensed by massive galaxies is also theoretically high (see Refs.~\cite{Oguri:2019fix,Liao:2022gde} for strongly lensed transient reviews).
In the following simulation of lensed FRB events, we calculate the time-delay distances with $2.15\%$ relative errors for them. 

\begin{figure*}[!ht]
\includegraphics[width=0.95\linewidth,angle=0]{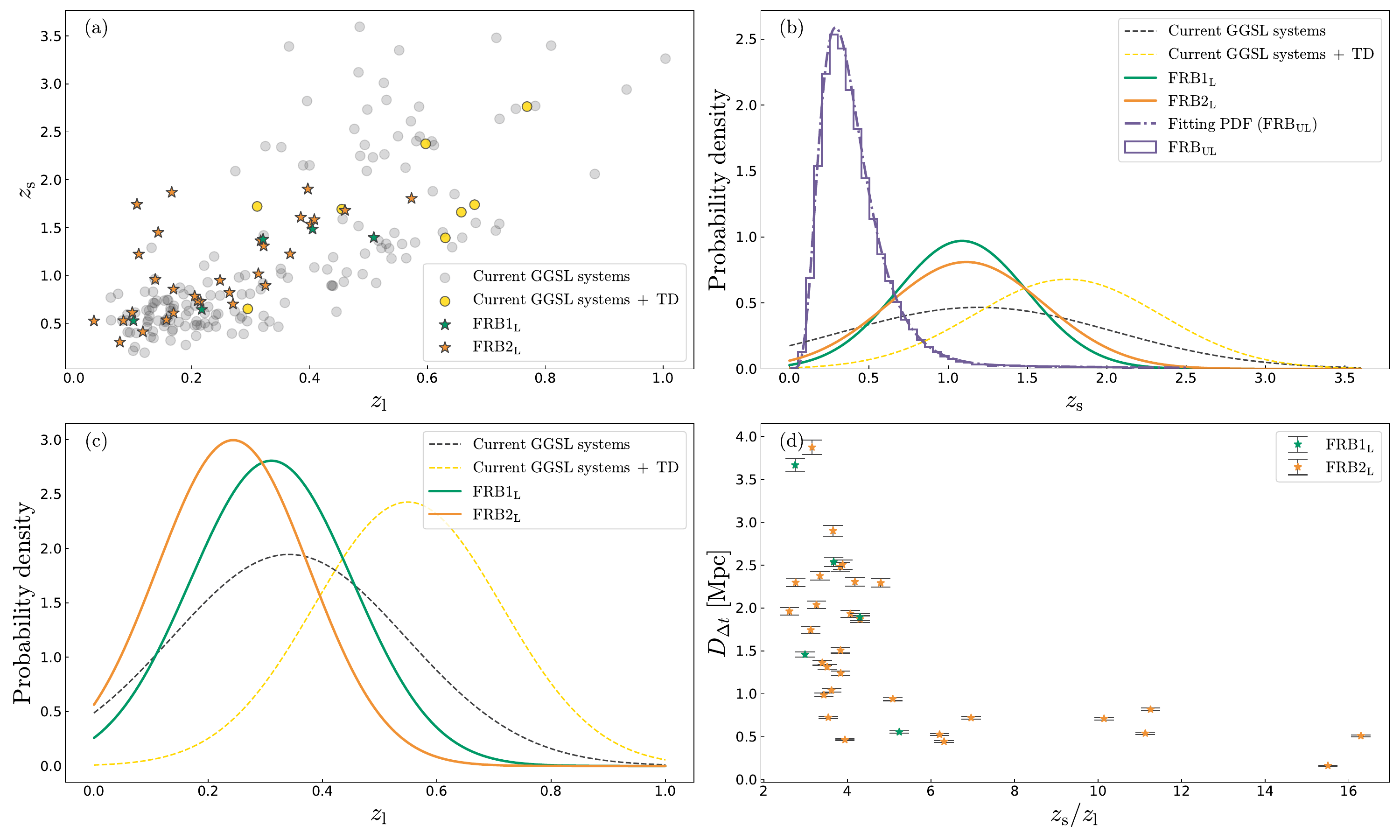}
\vspace{-0.3cm}
\caption{Simulated FRB data from a $5$-year CASM observation.
Panel (a) shows the distributions of lens redshifts ($z_{\rm l}$) versus source redshifts ($z_{\rm s}$), with grey circles indicating current GGSL systems from Ref.~\cite{Chen_2019} and yellow circles representing GGSL systems with TD measurements from Ref.~\cite{Denzel:2020zuq}.
The green and orange stars denote simulated GGSL FRB data for scenarios with $5$ (normal) and $40$ (optimistic) sample sizes, labeled as $\rm FRB1_{\rm L}$ and $\rm FRB2_{\rm L}$, respectively, which are derived from the simulated sample of GGSL systems from Ref.~\cite{Collett:2015roa}.
Note that $\rm FRB2_{\rm L}$ includes $\rm FRB1_{\rm L}$. 
Panels (b) and (c) display the normalized redshift distributions for $z_{\rm s}$ and $z_{\rm l}$, respectively. The lines of different colors and styles correspond to different datasets.
The histogram of host galaxy redshift of unlensed FRBs ($\rm FRB_{\rm UL}$) is also illustrated in panel (b), with the normalized redshift distribution plotted as the purple line to show the lognormal+Cauchy fitting function (see Eq.~(\ref{eq:pbimodal})). 
Panel (d) illustrates the simulated TD distance data. Note that the fiducial model using the fiducial flat dynamical dark energy model with $w$ = constant.}\label{data:lensed}
\end{figure*}

\subsection{DM measurement from unlensed FRBs} \label{dm}
We generate the DMs of unlensed FRB samples using the DM model in Ref.~\cite{Zhang:2023gye}. The observed DM, $\rm DM_{\rm obs}$, is a measure of the number density of free electrons $n_{\rm e}$ weighted by $(1+z)^{-1}$, along the path $l$ to the FRB: ${\rm DM}_{{\rm obs}}=\int n_{{\rm e}} dl /(1+z)$. This value can be determined by the captive signal with the time delay between the highest frequency and the lowest frequency. Physically, $\rm DM_{\rm obs}$ is usually divided to four components: two from the Milky Way, i.e., one from the interstellar medium (ISM) and a second from its halo; and two extragalactic ones, the IGM and the FRB host galaxy:
\begin{equation}
	\rm{DM}_{\rm{obs}}=\rm{DM}_{\rm{MW}}+\rm{DM}_{\rm{E}}. \label{eq:dmobs5}
\end{equation}

For the DM contribution within the Milky Way $\rm DM_{\rm MW}=\rm DM_{\rm MW,ISM}+\rm DM_{\rm MW,halo}$, 
$\rm DM_{\rm MW,ISM}$ can be obtained using the typical electron density models of the Milky Way, i.e., the NE2001 \cite{Cordes:2002wz} and YMW16 \cite{yao2017new} models. The calculation is related to the FRBs' Galactic coordinates.
$\rm DM_{\rm MW,halo}$ is in the range of [$30$, $80$] \pccc \cite{Dolag:2014bca,10.1093/mnras/stz261}. 
In this study, we use the NE2001 model to calculate \(\rm DM_{\rm MW, ISM}\) and assume a normal distribution to model $\rm DM_{\rm MW,halo}$ as $\mathrm{DM}_{\mathrm{MW,halo}} \sim \mathcal{N}(55, 25^2)$ (in units of \pccc) \cite{Wu:2021jyk}. 

On the other hand, the extragalactic contribution,
$\rm DM_{\rm E}=\rm DM_{\rm IGM}+\rm DM_{\rm host}$, is typically the dominant part in $\rm DM_{\rm obs}$. $\rm{DM}_{\rm{IGM}}$ is closely related to cosmology, and the Macquart relation gives its averaged value \cite{Macquart:2020lln},
\begin{equation}\label{aveDM}
	\langle\DM_{\rm IGM}(z)\rangle=\frac{3c \Omega_{\rm b} H_0^{2}}{8\pi G m_{\rm p}H_0}\int_0^z\frac{\chi(z')f_{\rm{IGM}}(z')(1+z')}{E(z')}dz',
\end{equation}
where $G$ is the gravitational constant, $m_{\rm{p}}$ is the mass of a proton, and $\chi(z)$ represents the number of free electrons per baryon, i.e., $\chi_e(z)=\frac{3}{4} \chi_{e, \rm{H}}(z)+\frac{1}{8} \chi_{e, \rm{He}}(z)$, where $\chi_{e, \rm{H}}$ and $\chi_{e, \rm{He}}$ are ionization fractions for hydrogen and helium, respectively. We take $\chi_{\rm{e,H}}=\chi_{\rm{e,He}}=1$, assuming that both hydrogen and helium are fully ionized at $z<3$. 
$f_{\rm{IGM}}(z)$ is the baryon fraction in the diffuse IGM evolving with redshift. It is suggested that $f_{\rm{IGM}} \simeq 0.9$ at $z \geq 1.5$ \cite{Meiksin:2007rz} and $f_{\rm{IGM}} \simeq 0.82$ at $z \leq 0.4$ \cite{Shull:2011aa} (see Refs.~\cite{Li:2019klc,Wei:2019uhh,Li:2020qei,Dai:2021czy,Wang:2022ami,Lin:2023opv,Lemos:2023eaq} for other studies constraining $f_{\rm IGM}$). We adopt a moderate value of $f_{\rm{IGM}} \simeq 0.83$ for the redshift range considered in our mock samples.
More importantly, $E(z)$, the dimensionless Hubble parameter, is directly related to cosmological parameters, which will be discussed further in sect.~\ref{cos:models}.

Due to large fluctuations in the IGM, the actual value of $\rm DM_{\rm IGM}$ varies significantly around the mean value $\langle{\rm{DM}}_{{{\rm{IGM}}}}(z)\rangle$. 
The variation is mainly attributed to the galactic feedback \cite{Walters:2017afr}. 
The probability distribution function (PDF) of $\rm DM_{\rm IGM}$ has been derived from numerical simulations of the IGM \cite{mcquinn2013locating} and galaxy halos \cite{10.1093/mnras/stz261}. 
Based on cosmological principles, the impact of compact halo contribution on large scales is insignificant, and we assume that the distribution of $\rm DM_{\rm IGM}$ follows a Gaussian distribution, with $\sigma_{\rm IGM}$ scaling with redshift in a power-law form as \cite{Qiang:2021bwb}: 
\begin{equation}
	\sigma_{\rm{IGM}}(z)=173.8z^{0.4}~\rm{pc}~\rm{cm}^{-3}.
\end{equation}

The contribution from host galaxy, $\rm DM_{\rm host}$, is difficult for modeling due to its strong dependence on the type of galaxy and local environment. Ref.~\cite{Macquart:2020lln} proposed a lognormal PDF with an asymmetric long tail allowing for high $\rm DM_{\rm host}$ values (e.g., that of FRB 20190520B \cite{Niu:2021bnl}), which fits well with the results from the IllustrisTNG simulation \cite{Zhang:2020mgq}.
However, recent simulations suggest that $\rm DM_{\rm host}$ may deviate from this distribution \cite{Beniamini:2020ane,Orr:2024ucd}. So the real distribution remains uncertain, which needs larger FRB samples in future studies. We adopt a simplified physical scenario, assuming that the distribution of $\rm DM_{\rm host}$ also follows a Gaussian distribution with a standard deviation of $\sigma_{\rm{host}} = 30 ~{\rm {pc~cm^{-3}}}$ \cite{Li:2019klc}, which is expected to be realized in the high-statistics era.

Overall, $\rm{DM}_{\rm{IGM}}$ is available for a localized FRB with $\rm{DM}_{\rm{obs}}$, $\rm{DM}_{\rm{MW}}$ and $\rm{DM}_{\rm{host}}$ determined.
The observational uncertainty is negligible compared to other errors and can be ignored. Consequently, if these parameters are treated properly, the total uncertainty of $\rm{DM}_{\rm{IGM}}$ is determined by 
\begin{equation}\label{eq:sigmadm}
	\sigma_{\rm{DM}_{\rm{IGM}}}=\left[\sigma_{\rm MW}^{2}+\sigma_{\rm IGM}^{2}
	+\left(\frac{\sigma_{\rm host}}{1+z}\right)^{2} \right]^{1/2},
\end{equation}
where the uncertainty of $\rm{DM}_{\rm{MW}}$, i.e., $\sigma_{\rm MW}$, averages about $10~{\rm {pc~cm^{-3}}}$ for the pulses from high Galactic latitude $\left(|b|>10^{\circ}\right)$. The factor $(1+ z)$ accounts
for cosmological time dilation for a source at redshift $z$.

For the redshift distribution of the mock unlensed FRB data, we have fitted the distribution in Ref.~\cite{Connor:2022bwl} with a lognormal$+$Cauchy PDF,
\begin{equation}
	N(z)=\frac{1}{\sqrt{2\pi}\sigma z} \exp \left[-\frac{\left(\lg z-\mu \right)^2}{2\sigma^2}\right]+\frac{\gamma_0}{\pi((x-x_{0})^2 + 1)}\label{eq:pbimodal},
\end{equation}
where parameters $\sigma$, $\mu$, $\gamma_0$, and $x_0$ are fitted to $0.45$, $-0.02$, $0.05$, 
and $1.60$, respectively. 
This is the assumed redshift distribution for the CASM, which is derived by scaling the redshift distribution from the latest CHIME catalog \cite{CHIMEFRB:2021srp}.

\subsection{Simulation of TD and DM measurements from CASM's FRBs}

\begin{table*}[!t]
	\centering
	\caption{Main properties of CHIME, BURSTT-256, and the hypothesized CASM in this work, corresponding to current, upcoming, and future-concept FRB observatories. Notably, the detection rate of CASM is 25 times higher than that of CHIME due to its 25 times larger FoV.}
	\label{tab:BURSTT}
	\renewcommand{\arraystretch}{1.5}
	\setlength{\tabcolsep}{28pt}
	\begin{tabular}{cccc}
		\hline \hline
		Specification & CHIME \cite{bandura2014canadian} & BURSTT-256 \cite{Lin:2022wgp} & CASM \cite{Connor:2022bwl}  \\
		\hline 
		Number of antennas & $1,024$ & $256$ & $\sim 25,000$ \\
		Frequency range (MHz) & $400$--$800$  & $300$--$800$ & $400$--$800$ \\
		Effective area ($\rm m^{2}$)  & $8,000$ & $40$--$200$ & $8,000$ \\
		FoV ($\rm deg^{2}$) & $200$ & 10,000 & 5,000 \\
		SEFD (Jy) & $50$ & $\sim 5,000$ & $50$   \\
		Localization accuracy ($''$) & $\sim 600$ & $<1$ & $<1$\\
		Detection event rate ($\rm yr^{-1}$) & $500$--$1,000$ & $\sim 100$ & $12,500$--$25,000$\\
		\hline \hline
	\end{tabular}
\end{table*}

We briefly introduce the CASM survey and then the simulation of both the TD distance and DM measurements from CASM’s FRBs. In this work, we focus on CASM survey, and do not consider instruments like SKA, which has a much smaller instantaneous FoV of only $\sim 20 ~\rm{deg}^{2}$.

So far, no gravitational lensing has been firmly identified in FRB signals (nevertheless, FRB 20190308C was recently reported as a plausible candidate with a significance of $3.4\sigma$ and requires further investigation \cite{Chang:2024hxi}).
This is likely due to missed detections, lensing below telescope sensitivity, or simply not being lensed within the observation period. To settle these issues, continuous coverage of a significant fraction of the sky may be the optimal way of finding strongly lensed FRBs \cite{Wucknitz:2020spz}.
The CASM facility can observe a unique all-sky collecting area, which will play a critical role in the blind FRB search. This is because surveys of transient events like FRBs benefit from a large FoV, as the number of detections increases proportionally with the FoV, while detecting persistent sources depends on sensitivity, which increases with the square root of the FoV \cite{Luo:2024zvw}.
An explicit example of CASM is the upcoming Bustling Universe Radio Survey Telescope for Taiwan (BURSTT) \cite{Lin:2022wgp} project\footnote{\url{https://www.burstt.org/}}. BURSTT is the first telescope dedicated for a complete census of nearby FRBs with a long time window, which allows for monitoring of FRBs for repetition and counterpart identification.
These would be clues to understanding the origins of FRBs, and there have been many efforts for answering whether all FRBs repeat \cite{Chen:2021jpq,Luo:2022smj,Zhu-Ge:2022nkz,Kirsten:2023eqd,Sun:2024huw} and what their counterparts are \cite{CHIMEFRB:2020abu,bochenek2020fast,Lin:2020mpw,li2021hxmt,LIGOScientific:2022jpr,Moroianu:2022ccs,Qi-lin:2024ypx}.

Due to its unique fisheye design, BURSTT has an extremely large instantaneous FoV of $ \sim 10,000~\rm deg^{2}$.
With VLBI outrigger stations, BURSTT can achieve a sub-acrsecond localization for identifying the unambiguous host galaxy.
More importantly, BURSTT has been considered to search for lensed FRBs with short time delay (less than $\sim$ ms) \cite{Ho:2023feo}.
Meanwhile, its wide FoV enables long-term, high-cadence monitoring of a large sky area, which can technically detect lensing delays up to the survey duration, and avoids missing potentially lensed FRBs with long time delays from days to months. This suggests that CASM experiments are particularly superior in detecting strongly lensed FRBs.

However, BURSTT is initially designed to detect hundreds of bright and nearby FRBs at $z \sim 0.03$ per year.
To detect strongly lensed FRBs usually at higher redshifts, a high sensitivity is required for blind searches, which can be achieved by using a dense aperture array with substantial small antennas.
BURSTT is built with 256 antennas (BURSTT-256) and is planned to expand to 2048 antennas (BURSTT-2048). The SEFD of BURSTT-256 is $\sim 5,000$ Jy, and that of BURSTT-2048 is $\sim$ 600 Jy, but both are still higher than CHIME's SEFD of $50$ Jy.
In principle, methods to improve sensitivity include reducing system temperature, increasing the number of antennas, and utilizing a compact phased array technology for beamforming \cite{Luo:2024zvw}. If the sensitivity matches that of CHIME, the number of detected FRBs would dramatically increase.
Ref.~\cite{Connor:2022bwl} assumed a future CASM survey with a large FoV of $ \sim 5,000~\rm deg^{2}$ and an SEFD of $50$ Jy observing in the $400$--$800$ MHz band, which is equivalent to a CHIME/FRB survey but with $25$ times of the sky coverage. They predicted that such a survey could potentially detect 50,000--100,000 FRBs including $5$--$40$ lensed events, during a 5-year observation with an $80\%$ duty cycle (see Table 2 in Ref.~\cite{Connor:2022bwl}).
The expected lensing probability for each CASM's FRB is $\sim \mathcal{O}(10^{-5})$--$\mathcal{O}(10^{-4})$, which is estimated using a subset of one-off events from the first CHIME/FRB Catalog (see Appendix \ref{Appendix} for a brief summary about the lensing event rate estimation). Despite the low probability of detecting galaxy-galaxy lensing,
the large number of FRBs detected with the CASM enhances the likelihood of capturing lensed events, which is benefited by the large sky coverage and long-term monitoring.
If these FRBs can be well localized, they  would be valuable for studying FRB cosmology, which requires a large sample of localized FRBs as well as potential lensed events.
This is also what future BURSTT science pursues, focusing on key topics in cosmology and fundamental physics.
We assumed that the CASM can also
localize FRBs to the sub-acrsecond resolution like BURSTT, so the redshifts of the detected FRBs can be measured.
The main properties of the CASM are summarized in the last column of Table~\ref{tab:BURSTT}, together with CHIME and BURSTT-256 for comparison.

In this work, we simulate the FRB data observed by the CASM survey (with a CHIME/FRB SEFD), using the methods for simulating TD measurement in sect.~\ref{td} and DM measurement in sect.~\ref{dm}.
Based on the event rate estimation in Ref.~\cite{Connor:2022bwl}, the lower-bound case predicts $5$ lensed FRBs assuming a total detection of 50,000 FRBs and no magnification bias, and the upper-bound case estimates $40$ lensed FRBs assuming 100,000 detected FRBs and strong magnification bias (see Appendix \ref{Appendix} for more details).
Since we focus on what role the lensed bursts can play in large samples, we assume a total detection of 100,000 FRBs for convenience.
Thus, we define two scenarios: a conservative scenario with $5$ lensed FRBs (labeled as $\rm FRB1_{\rm L}$) and an optimistic scenario with $40$ lensed FRBs (labeled as $\rm FRB2_{\rm L}$), both within a sample of 100,000 unlensed FRBs (labeled as $\rm FRB_{\rm UL}$).
The mock lensed FRB data, which are derived from the mock strong lens sample in Ref.~\cite{Collett:2015roa}, are shown in Fig.~\ref{data:lensed}.
In addition, the histogram of simulated unlensed FRBs and the fitting PDF (see Eq.~(\ref{eq:pbimodal})) are illustrated in purple in Fig.~\ref{data:lensed}(b).
The redshift distributions of the mock lensed FRBs are consistent with that of the unlensed FRBs and significantly overlap with both the currently detected GGSL systems from Ref.~\cite{Chen_2019} and those with TD measurements from Ref.~\cite{Denzel:2020zuq}.

\subsection{Other cosmological data}
In addition to mock FRB data, we incorporate complementary cosmological datasets, including the actual CMB, BAO and SNe data, as well as mock GW datasets.

For the mock data, 
we employ the GW {\it standard siren} data from Ref.~\cite{Zhang:2023gye}, which is generated based on the third-generation ground-based GW detector, the Einstein Telescope (ET) \cite{Punturo_2010}.
The GW standard siren method \cite{Schutz:1986gp,Holz:2005df} is an emerging probe of the late universe, which gives $\sim 14\%$ precision in measuring $H_0$ through the only multi-messenger observations of GW170817
\cite{LIGOScientific:2017adf}. Recently, this method has widely informed forecasts of the cosmological parameter estimations based on the future standard sirens detected by ground-based \cite{Zhao:2010sz,Chen:2017rfc,Wang:2018lun,Zhang:2018byx,Du:2018tia,Chang:2019xcb,Zhang:2019ple,Jin:2020hmc,Jin:2021pcv,Cao:2021zpf,Wang:2022rvf,Jin:2022tdf,Jin:2022qnj,Li:2023gtu,Han:2023exn,Jin:2023tou,Feng:2024lzh,Dong:2024bvw,Zheng:2024mbo,Fu:2024cjj,Feng:2024mfx}, space-borne \cite{Cutler:2009qv,Cai:2017aea,Wang:2019tto,Zhao:2019gyk,Wang:2021srv,guo2022standard,Song:2022siz,Jin:2023sfc,zhaowenscpma,zhuzonghongscpma} GW observatories, and pulsar timing arrays projects \cite{Yan:2019sbx,Wang:2022oou,Xiao:2024nmi}. These efforts detect GWs across various frequency bands from nanohertz to several hundred hertz (see Refs.~\cite{Cai:2017cbj,Zhang:2019ylr,Bian:2021ini} for reviews).
The absolute distance to a GW source can be determined by analyzing the GW waveform, where the amplitude in frequency domain is approximately inversely proportional to the luminosity distance $D_{\rm{L}}$.
We consider only the binary neutron star (BNS) merger events, which could provide electromagnetic (EM) counterparts carrying redshift information.
Then the established $D_{\rm{L}}$--$z$ relation can be used to study cosmology as
\begin{equation}\label{eq:dl}
	D_{\rm{L}}(z)=(1+z) \int_{0}^{z} \frac{d z^{\prime}}{H(z^{\prime})},
\end{equation}
which is referred to as bright sirens. We only consider bright siren data in this work. The error in the measurement of $D_{\rm L}$ is calculated as
\begin{equation}\label{eq:sigmadl}
	\sigma_{D_{\rm L}} = \left[ (\sigma_{D_{\rm L}}^{\rm inst})^2 + (\sigma_{D_{\rm L}}^{\rm lens})^2 + (\sigma_{D_{\rm L}}^{\rm pv})^2 \right]^{1/2},
\end{equation}
where $\sigma_{D_{\rm L}}^{\rm inst}$, $\sigma_{D_{\rm L}}^{\rm lens}$, and $\sigma_{D_{\rm L}}^{\rm pv}$ are the instrumental, weak lensing, and peculiar velocity errors derived from Refs.~\cite{Nishizawa_2011}, \cite{Hirata_2010}, and \cite{Gordon_2007}, respectively. 
The ET is anticipated to detect $1,000$ bright sirens from BNS mergers at $z\lesssim5$ during a 10-year observation period \cite{Zhang:2019loq}. 
For more details on this simulation, readers can refer to Ref.~\cite{Zhang:2023gye}.

In addition to mock data, we also utilize actual mainstream cosmological data, including the CMB, BAO, and SNe data.
For the CMB data, we employ the ``Planck distance priors'' from the Planck 2018 observation \cite{Chen_2019}, and we use the BAO measurements from 6dFGS at $z_{\rm eff} = 0.106$ \cite{Beutler_2011}, SDSS-MGS at $z_{\rm eff} = 0.15$ \cite{Ross_2015}, and Data
Release $12$ of Baryon Oscillation Spectroscopic Survey (BOSS-DR12) at $z_{\rm eff} = 0.38$, $0.51$, and $0.61$ \cite{Alam_2017}. 
For the SNe data, we use the sample from the Pantheon compilation with $1048$ supernovae data \cite{Scolnic_2018}.

We use the data combination CMB+BAO+SNe (abbreviated as ``CBS'') to constrain the fiducial cosmological models, and the obtained best-fit cosmological parameters are used to generate the central values of DMs and TDs in the simulated FRB data.

\subsection{Cosmological parameter estimation} \label{cos:models}
\begin{table*}[!t]
	\caption{Absolute ($1\sigma$) and relative errors on cosmological parameters of interest in the $w$CDM and $w_{0}w_{a}$CDM models using the single dataset from early- or late-universe observation, i.e., CMB, $\rm FRB_{\rm UL}$, $\rm FRB1_{\rm L}$, and $\rm FRB2_{\rm L}$. Note that the mock FRB data are derived from a 5-year observation of BURSTT, including $\rm FRB_{\rm UL}$ with an unlensed event number of 100,000, as well as $\rm FRB1_{\rm L}$ and $\rm FRB2_{\rm L}$ with lensed event numbers of $5$ and $40$, respectively. Here, $H_0$ is in units of km s$^{-1}$ Mpc$^{-1}$.}
	\label{tab:3}
	\setlength{\tabcolsep}{9mm}
	\renewcommand{\arraystretch}{1.5}
	\begin{center}{\centerline{
				\begin{tabular}{cccccc}
					\hline \hline
					Model & Error 
					& CMB  
					& $\rm FRB_{\rm UL}$ & $\rm FRB1_{\rm L}$ & $\rm FRB2_{\rm L}$ \\
					\hline
					\multirow{4}{*}{$w$CDM}
					& $\sigma(H_{0})$   
					& 6.20  
					& ---   & 1.35  & 0.52  \\
					& $\varepsilon(H_{0})$   
					& $9.0\%$ 
					& --- & $2.0\%$ & $0.8\%$ \\ 
					& $\sigma(w)$   
					& 0.200     
					& 0.052  & --- & 0.245 \\
					& $\varepsilon(w)$   
					& $19.2\%$ 
					& $5.1\%$ & --- & $19.6\%$ \\ 
					\hline
					\multirow{5}{*}{$w_{0}w_{a}$CDM}
					& $\sigma(H_{0})$   
					& 7.60 
					& ---   & 1.65 & 0.70 \\ 
					& $\varepsilon(H_{0})$   
					& $11.0\%$ 
					& ---   & $2.4\%$ & $1.0\%$ \\ 
					& $\sigma(w_{0})$   
					& 0.450 
					& 0.105  & ---  & 0.835 \\
					& $\varepsilon(w_{0})$   
					& $54.0\%$ 
					& $11.0\%$ & --- & $68.0\%$ \\ 
					& $\sigma(w_{a})$  
					& ---  
					& 0.85  & ---  & 1.65  \\
					\hline \hline
		\end{tabular}}}
	\end{center}
\end{table*}
We provide an overview of the adopted dark energy models and methods for cosmological parameter estimation.

The dark energy models considered in this work include flat $\Lambda$CDM, $w$CDM, and $w_{0}w_{a}$CDM models.
The EoS parameter for dark energy, $w(z)$, is defined as the ratio of its pressure $p_{\rm{de}}(z)$ to density $\rho_{\rm{de}}(z)$ at redshift $z$. 
It helps describe the dimensionless Hubble parameter $E(z)$, which can be formulated using the Friedmann equation for a given cosmological model.
In a spatially-flat universe, the dimensionless Hubble parameter $E(z)=H(z)/H_0$ is expressed as
\begin{equation}\label{eq
	} E^2(z) = \Omega_{\rm m}(1 + z)^3 + (1 - \Omega_{\rm m}) \exp\left[3 \int_0^z \frac{1 + w(z')}{1 + z'} dz'\right], \end{equation}
where $\Omega_{\rm m}$ represents the present-day matter density parameter.

The simplest and most widely accepted dark energy model is the $\Lambda$CDM model  with the vacuum energy EoS $w = -1$. The $w$CDM model, on the other hand, assumes a constant EoS $w$, representing the simplest scenario for dynamical dark energy. Finally, the $w_{0}w_{a}$CDM model describes an evolving EoS as $w(z) = w_{0} + w_{a}/(1+z)$, where $w_{0}$ and $w_a$ are the two parameters that characterize the evolution.

The cosmological parameters $\boldsymbol{\xi}$ we sample include $\Omega_{\rm m}$, $\Omega_{\rm b}h^{2}$, $w$, $w_0$, $w_a$, and $H_0$, and we take flat priors within ranges of $[0, 1]$, $[0, 0.05]$, $[-2, 1]$, $[-5, 1]$, $[-3, 3]$, and $[0, 100]$ $\rm km\ s^{-1}\ Mpc^{-1}$, respectively. 
{Note that for the baryonic parameter, we have the options of $\Omega_{\rm b}$, $\Omega_{\rm b}h$, and $\Omega_{\rm b}h^{2}$ (see Eq.~(\ref{aveDM})). Since baryons are not the focus of this work, we only adopt $\Omega_{\rm b}h^{2}$, which is directly constrained by the CMB data.}
They are optimized via maximization of the joint likelihood function 
\begin{equation}
	\mathcal{L} \propto \rm{e}^{-\chi^2/2},
\end{equation}
where the $\chi^2$ function for lensed FRBs, unlensed FRBs, and GWs are defined as
\begin{equation}
	\chi_{\mathrm{TD}}^2(\boldsymbol{\xi})=\sum_{i=1}^{N_{\rm FRB_{\rm L}}}\left(\frac{D_{\Delta t, i}^{\mathrm{th}}(\boldsymbol{\xi})-D_{\Delta t, i}^{\mathrm{obs}}}{\sigma_{D_{\Delta t}}(z_{i})}\right)^2,
\end{equation}

\begin{equation}
	\chi_{\rm DM}^{2}(\boldsymbol{\xi})=\sum_{i=1}^{N_{\rm FRB_{\rm UL}}}\left(\frac{{\langle\rm{DM}_{\rm{IGM}}^{\rm{th}}}(z_{i}; {\boldsymbol{\xi}})\rangle-{\rm{DM}_{\rm{IGM}}^{\rm{obs}}}(z_{i})}{\sigma_{\rm{DM}_{\rm{IGM}}}(z_{i})}\right)^{2},
\end{equation}
and 
\begin{equation}
	\chi_{\rm{GW}}^{2}(\boldsymbol{\xi})=\sum_{i=1}^{N_{\rm GW}}\left(\frac{D_{\rm{L}}^{\rm{th}}(z_{i}; {\boldsymbol{\xi}})-D_{\rm{L}}^{\rm{obs}}(z_{i})}{\sigma_{D_{\rm L}}(z_{i})}\right)^{2},
\end{equation}
respectively.
$D_{\Delta t,i}^{\mathrm{obs}}$, $\rm{DM}_{\rm{IGM}}^{\rm {obs}}$, and $D_{\rm{L}}^{\rm{obs}}$ are the observable values, and
$D_{\Delta t, i}^{\mathrm{th}}(\boldsymbol{\xi})$, ${\rm DM}_{\rm{IGM}}^{\rm{th}}(z_i; \boldsymbol{\xi})$, and $D_{\rm{L}}^{\rm{th}}(z_{i}; {\boldsymbol{\xi}})$
are theoretical values of $D_{\Delta t}$, $\rm{DM}_{\rm{IGM}}$ and $D_{\rm{L}}$ at $z_i$ calculated by Eqs.~(\ref{eq:deltat}), (\ref{aveDM}), and (\ref{eq:dl}), respectively.
$\sigma_{D_{\Delta t}}(z_{i})$, $\sigma_{\rm{DM}_{\rm{IGM}}}$($z_{i}$), and $\sigma_{d_{\rm L}}(z_{i})$ represent the related uncertainties calculated by Eqs.~(\ref{eq:sigmaDdeltat}), (\ref{eq:sigmadm}), and (\ref{eq:sigmadl}).

We derive the posterior probability distributions through the Markov chain Monte Carlo analysis (MCMC) ensemble sampler {\tt emcee} \cite{foreman2013emcee}, and use {\tt GetDist}\footnote{\url{https://github.com/cmbant/getdist/}} for plotting the posterior distributions of the cosmological parameters.

\section{Results and discussion}\label{sec3}
\begin{table*}[!htbp]
\caption{Same as Table~\ref{tab:3} but using the combined datasets from both early- and late-universe observations, i.e., CBS, CMB+$\rm FRB_{\rm UL}$, CMB+$\rm FRB1_{\rm L}$, and CMB+$\rm FRB2_{\rm L}$ data. Here, CBS stands for CMB+BAO+SNe and $H_0$ is in units of km s$^{-1}$ Mpc$^{-1}$.}
\label{tab:4}
\setlength{\tabcolsep}{7mm}
\renewcommand{\arraystretch}{1.5}
\begin{center}{\centerline{
\begin{tabular}{cccccc}
\hline \hline
Model & Error & CBS & CMB+$\rm FRB_{\rm UL}$ & CMB+$\rm FRB1_{\rm L}$ & CMB+$\rm FRB2_{\rm L}$ \\
\hline
\multirow{4}{*}{$w$CDM}
& $\sigma(H_{0})$   & 0.83 & 0.35 & 0.79 & 0.28  \\
& $\varepsilon(H_{0})$   & $1.2\%$ & $0.5\%$ & $1.2\%$ & $0.4\%$  \\
& $\sigma(w)$   & 0.034 & 0.019 & 0.036 & 0.026 \\
& $\varepsilon(w)$   & $3.4\%$ & $1.9\%$ & $3.5\%$ & $2.6\%$  \\ 
\hline
\multirow{5}{*}{$w_{0}w_{a}$CDM}
& $\sigma(H_{0})$   & 0.84 & 0.48 & 1.50 & 0.55   \\
& $\varepsilon(H_{0})$   & $1.2\%$ & $0.7\%$ & $2.2\%$ & $0.8\%$  \\
& $\sigma(w_{0})$   & 0.084 & 0.068 & 0.280 & 0.110 \\
& $\varepsilon(w_{0})$   & $8.5\%$ & $6.8\%$ & $30.4\%$ & $11.1\%$  \\
& $\sigma(w_{a})$   & 0.32 & 0.21 & 0.90 & 0.35 \\
\hline \hline
\end{tabular}}}
\end{center}
\end{table*}

\begin{figure*}
\begin{center}
\includegraphics[width=0.95\linewidth,angle=0]{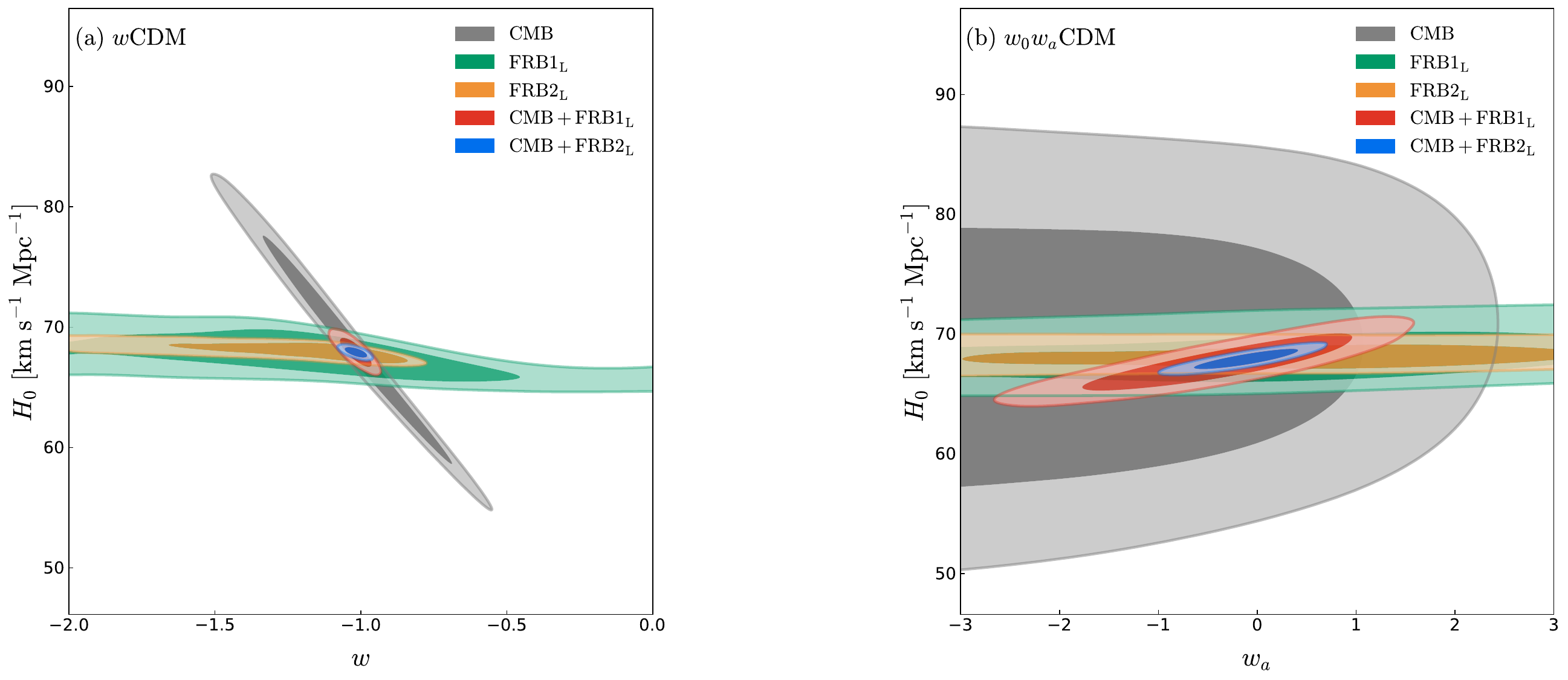}
\end{center}
\vspace{-0.3cm} 
\caption{Two-dimensional posterior distribution ($68.3\%$ and $95.4\%$ credible regions) in the $w$--$H_{0}$ plane for the $w$CDM model (a) and the $w_{a}$--$H_{0}$ plane for the $w_{0}w_{a}$CDM model (b), by using the CMB, FRB$1_{\rm L}$, FRB$2_{\rm L}$, CMB+FRB$1_{\rm L}$, and CMB+FRB$2_{\rm L}$ data.} \label{Fig:CMB+FRBL}
\end{figure*}
In this section, we employ the simulated FRB observation from the CASM to constrain two dynamical dark energy models, i.e., $w$CDM and $w_{0}w_{a}$CDM models. For comparison and combination, we also use the actual CMB, BAO, and SNe data, as well as the mock GW data.
The constraint results for key cosmological parameters, i.e., $H_0$ and the EoS of dark energy, are summarized in Tables~\ref{tab:3}--\ref{tab:6} and shown in Figs.~\ref{Fig:CMB+FRBL}--\ref{Fig:CBS+FRBUL+FRBL} using different datasets. 
Here, we use $\sigma(\xi)$ and $\varepsilon(\xi)=\sigma(\xi)/\xi$ to represent the absolute and relative errors of the parameter $\xi$, respectively.

In the following, we first report the results separately from unlensed and lensed events in sect.~\ref{ULL alone}, and their joint analyses with CMB in sect.~\ref{CMBLUL}. Then we present the results from combining the lensed and the unlensed FRBs into a full sample in sect.~\ref{FRBULL}. Finally, we discuss the combination of this full FRB dataset with CBS in sect.~\ref{CBS+FRBULL}.

\subsection{The lensed or unlensed FRB data} \label{ULL alone}
We first report the constraint results from the FRB dataset of either unlensed or lensed events, i.e., $\rm FRB_{\rm UL}$, $\rm FRB1_{\rm L}$, and $\rm FRB2_{\rm L}$ in Table~\ref{tab:3},
which represent a single probe from the late-universe observation.

When considering dynamical dark energy, the CMB data alone cannot give precise constraints, with $\varepsilon(H_{0})=9.0\%$ and $\varepsilon(w)=19.2\%$ in the $w$CDM model.
In contrast, the FRB datasets can provide more precise constraints. A large number of 100,000 FRB data can effectively constrain dark-energy EoS parameters.
Concretely, $\rm FRB_{\rm UL}$ gives an absolute error of $\sigma(w)=0.052$ in the $w$CDM model, and $\sigma(w_0)=0.053$ and $\sigma(w_a)=0.85$ in the $w_{0}w_{a}$CDM model. The constraints are all about $74\%$ better than those from the actual CMB data.
However, it cannot constrain $H_0$ due to the strong parameter degeneracy between $\Omega_{\rm b}h^{2}$ and $H_0$ in Eq.~(\ref{aveDM}).

On the other hand, if the lensed events are detected, they will provide very precise measurement on $H_0$.
We give some examples in the $w_{0}w_{a}$CDM model.
By using $5$ and $40$ lensed FRB data, FRB$1_{\rm L}$ and FRB$2_{\rm L}$ offer $H_0$ constraints with $2.4\%$ and $1.0\%$ precision, respectively, which are about $78\%$ and $92\%$ more precise than that of the CMB data, respectively. 
Nevertheless, using the lensed data is insufficient to effectively constrain dark energy, with the most precise constraint of $\sigma(w)=0.245$.

Overall, the future FRB observation with the CASM can effectively measure dark energy and the Hubble constant, by analyzing DM and TD measurements from localized FRB datasets, respectively. 

\subsection{Combination with the CMB data} \label{CMBLUL}
\begin{figure}[htbp]
\includegraphics[width=0.85\linewidth,angle=0]{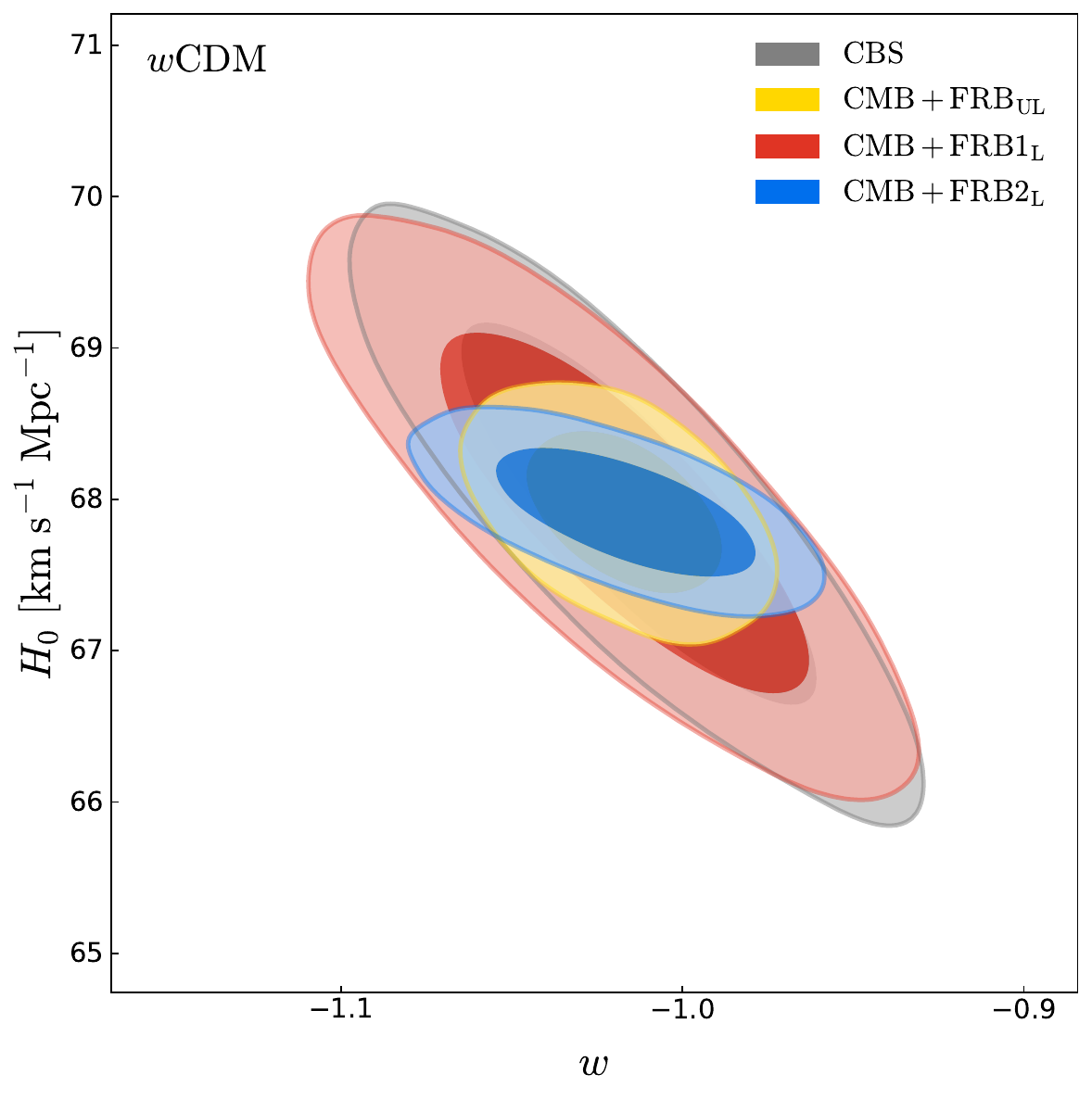}
\vspace{-0.3cm} 
\caption{Two-dimensional posterior distribution ($68.3\%$ and $95.4\%$ credible regions) in the $w$--$H_0$ plane for the $w$CDM model, by using the CBS, CMB+FRB$_{\rm UL}$, CMB+FRB$1_{\rm L}$, and CMB+FRB$2_{\rm L}$ data.} \label{Fig:compare_CMB+FRBL}
\end{figure}
Next, we report the results from combining the CMB data with FRB datasets of either unlensed or lensed events, i.e., CMB+$\rm FRB_{\rm UL}$, CMB+$\rm FRB1_{\rm L}$, and CMB+$\rm FRB2_{\rm L}$ in Tables~\ref{tab:4}, which represent multiple probes from both early- and late-universe observations.
We mainly discuss the results from the combined case of the lensed FRBs.

Current mainstream low-redshift observations, BAO and SNe, cannot independently constrain dark energy but can effectively break parameter degeneracies inherent in the CMB. By combining CMB with BAO+SNe, the CBS data offers greatly improved constraints: $\varepsilon(H_{0})=1.2\%$ and $\varepsilon(w)=3.4\%$.

Similarly, we find that when combining the CMB data, the lensed FRB data can precisely constrain both $H_0$ and dark-energy EoS parameters. The combination greatly improves the constraints over either single probe.
We give some examples in the $w$CDM model.
When compared to the CMB data alone, CMB+FRB$1_{\rm L}$ provides constraints in precision of $\varepsilon(H_{0})=1.2\%$ and $\varepsilon(w)=3.5\%$, which are improved by about $87\%$ and $82\%$, respectively.
Also, when compared to the FRB$1_{\rm L}$ and FRB$2_{\rm L}$ data alone, CMB+FRB$1_{\rm L}$ and CMB+FRB$2_{\rm L}$ improve the constraints on $H_0$ by about $40\%$ and $50\%$, respectively.
It is noteworthy that the lensed FRBs can effectively probe dark energy with the help of CMB.
Figs.~\ref{Fig:CMB+FRBL}(a) and~\ref{Fig:CMB+FRBL}(b) show 
the posterior contours in the $w$--$H_0$ and $w_{a}$--$H_0$ planes, respectively, using the CMB, FRB$1_{\rm L}$, FRB$2_{\rm L}$, CMB+FRB$1_{\rm L}$, and CMB+FRB$2_{\rm L}$ data. The results indicate that the lensed FRB data can well break the parameter degeneracy induced by CMB.

To assess the extent of this capability, we make some comparative analyses.
Previous studies~\cite{Liu:2019jka,Zhao:2020ole,Qiu:2021cww,Zhao:2022bpd} demonstrated that combining CMB with FRB data can effectively improve cosmological constraints.
Accordingly, we include the constraint contours of both CBS and CMB+$\rm FRB_{\rm UL}$ in Fig.~\ref{Fig:compare_CMB+FRBL} for comparison with those of CMB+FRB$1_{\rm L}$ and CMB+FRB$2_{\rm L}$ for the $w$CDM model. 

When compared to CBS, we can see that CMB+FRB$1_{\rm L}$ gives similar constraints on both $H_0$ and $w$, and CMB+FRB$2_{\rm L}$ provides even tighter measurements for these parameters. 
Note that for the $w_{0}w_{a}$CDM model, CMB+FRB$2_{\rm L}$ also offers constraints similar to CBS, with a better precision of $\varepsilon(H_{0})=0.8\%$, and slightly less precise results of $\varepsilon(w_{0})=11.1\%$ and $\sigma(w_{a})=0.35$.
These suggest that using $5$--$40$ lensed FRBs alone can effectively break the inherent parameter degeneracies in CMB, comparable to using the combination of BAO+SNe.

On the other hand, when compared to CMB+$\rm FRB_{\rm UL}$,
the constraints on $H_0$ from CMB+FRB$2_{\rm L}$ are more precise than those from CMB+$\rm FRB_{\rm UL}$ in Fig.~\ref{Fig:compare_CMB+FRBL}.
Specifically, CMB+FRB$2_{\rm L}$ achieves $20\%$ better precision of $\varepsilon(H_{0})=0.4\%$, and slightly worse precision of $\varepsilon(w)=2.6\%$ than CMB+FRB$_{\rm UL}$ (giving $\varepsilon(H_{0})=0.5\%$ and $\varepsilon(w)=1.9\%$).
They give similar constraint results.

Overall, we conclude that the effect of only $5$ lensed FRBs could be similar to that of BAO+SNe in breaking the CMB degeneracies, and $40$ lensed FRBs could even be comparable to 100,000 unlensed FRBs.

\subsection{The joint lensed and unlensed FRB data} \label{FRBULL}

\begin{table*}[!tbp]
	\caption{Same as Table~\ref{tab:3} but using the combined datasets from late-universe observations alone, i.e., $\rm FRB_{\rm UL}$+BAO+SNe, $\rm FRB_{\rm UL}$+GW, $\rm FRB_{\rm UL}+FRB1_{\rm L}$, and $\rm FRB_{\rm UL}+FRB2_{\rm L}$ data. Note that the mock GW data are derived from a 10-year observation of ET, with an expected event number of 1,000. Here, $H_0$ is in units of km s$^{-1}$ Mpc$^{-1}$.}
	\label{tab:5}
	\setlength{\tabcolsep}{5mm}
	\renewcommand{\arraystretch}{1.5}
	\begin{center}{\centerline{
				\begin{tabular}{cccccc}
					\hline \hline
					Model & Error 
					& $\rm FRB_{\rm UL}$+BAO+SNe  & $\rm FRB_{\rm UL}$+GW
					& $\rm FRB_{\rm UL}$+$\rm FRB1_{\rm L}$ & $\rm FRB_{\rm UL}$+$\rm FRB2_{\rm L}$ \\
					\hline
					\multirow{4}{*}{$w$CDM}
					& $\sigma(H_{0})$   
					& 2.50   & 0.42 
					& 0.71   & 0.30 \\
					& $\varepsilon(H_{0})$   
					& $1.2\%$ & $0.6\%$
					& $1.0\%$ & $0.4\%$ \\ 
					& $\sigma(w)$   
					& 0.045  & 0.049
					& 0.051   & 0.046 \\
					& $\varepsilon(w)$   
					& $4.4\%$ & $4.8\%$
					& $5.0\%$ & $4.5\%$ \\ 
					\hline
					\multirow{5}{*}{$w_{0}w_{a}$CDM}
					& $\sigma(H_{0})$   
					& 4.15    & 0.80 
					& 0.89    & 0.51 \\ 
					& $\varepsilon(H_{0})$   
					& $6.1\%$ & $1.2\%$
					& $1.3\%$ & $0.7\%$ \\ 
					& $\sigma(w_{0})$   
					& 0.068   & 0.095
					& 0.096   & 0.072 \\
					& $\varepsilon(w_{0})$   
					& $6.9\%$ & $9.8\%$
					& $9.9\%$ & $7.4\%$ \\ 
					& $\sigma(w_{a})$  
					& 0.61   & 0.80 
					& 0.79   & 0.70 \\
					\hline \hline
		\end{tabular}}}
	\end{center}
\end{table*}

\begin{figure*}[htbp]
\begin{center}
\includegraphics[width=0.95\linewidth,angle=0]{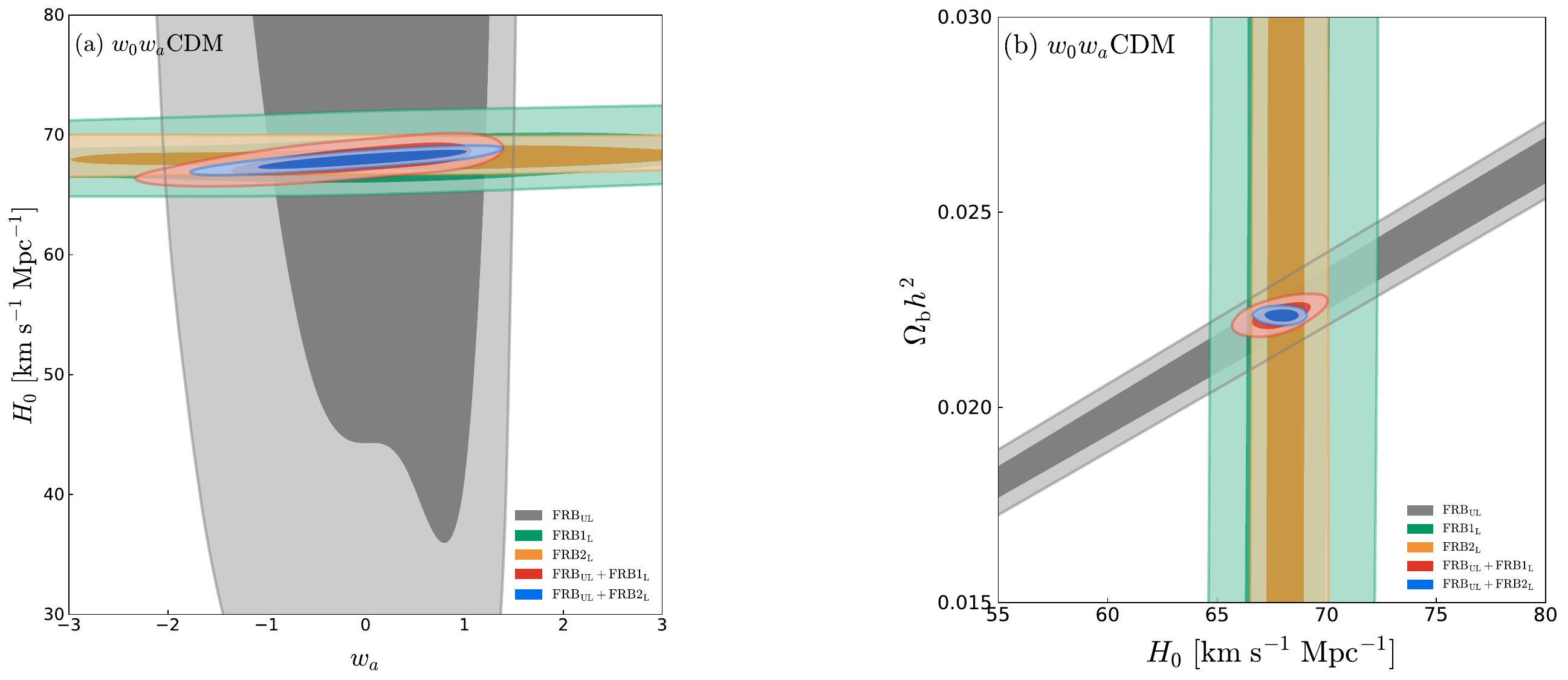}
\vspace{-0.3cm} 
\end{center}
\caption{Two-dimensional posterior distribution ($68.3\%$ and $95.4\%$ credible regions) in the $w_a$--$H_{0}$ plane (a) and the $H_0$--$\Omega_{\rm b}h^2$ plane (b) for the $w_{0}w_{a}$CDM model, by using different FRB samples, i.e., the $\rm FRB_{\rm UL}$, FRB$1_{\rm L}$, $\rm FRB2_{\rm L}$, $\rm FRB_{\rm UL}$+FRB$1_{\rm L}$, and $\rm FRB_{\rm UL}$+FRB$2_{\rm L}$ data.}
\label{Fig:FRBUL+FRBL}
\end{figure*}

\begin{figure*}[htbp]
	\begin{center}
		\includegraphics[width=0.95\linewidth,angle=0]{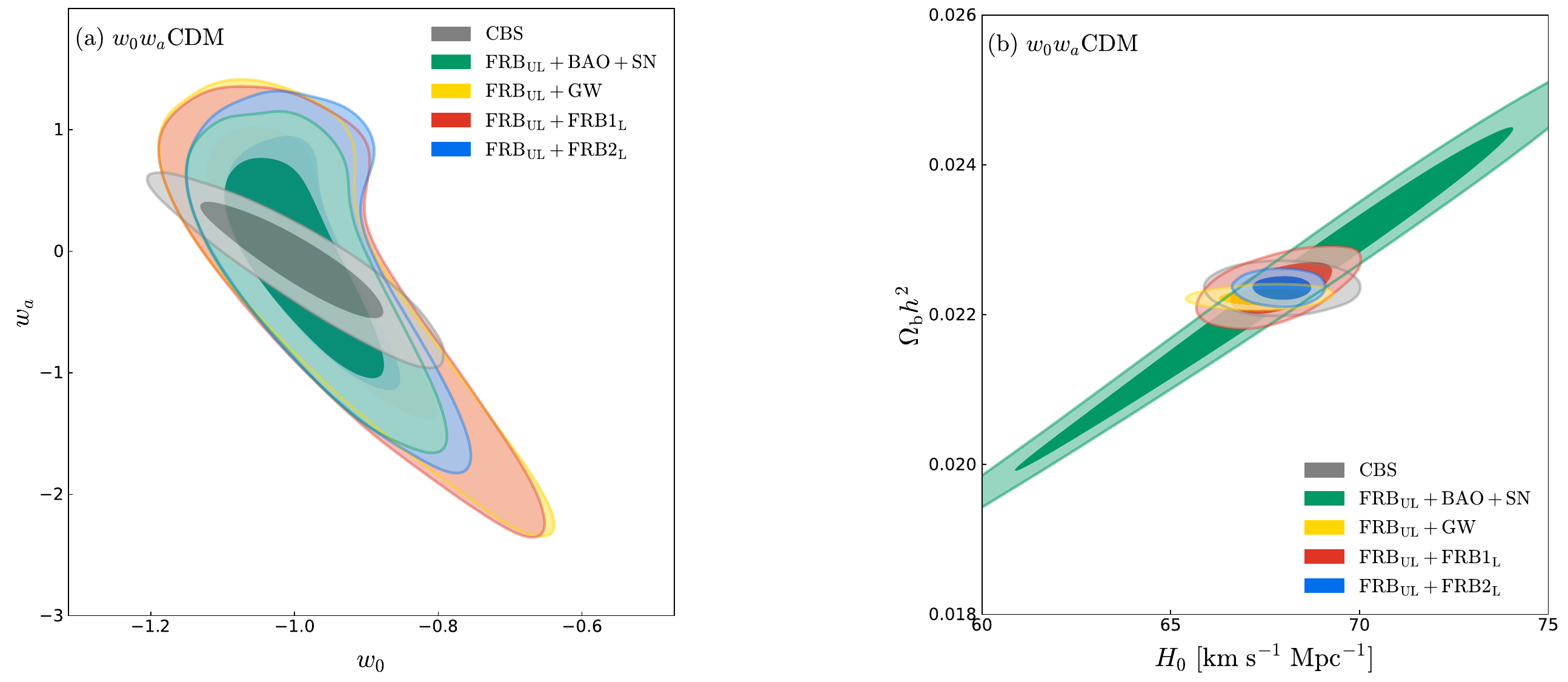}
		\vspace{-0.3cm} 
	\end{center}
	\caption{Two-dimensional posterior distribution ($68.3\%$ and $95.4\%$ credible regions) in the $w_0$--$w_a$ plane (a) and the $H_0$--$\Omega_{\rm b}h^2$ plane (b) for the $w_{0}w_{a}$CDM model, by using the CBS, $\rm FRB_{\rm UL}$+BAO+SNe, $\rm FRB_{\rm UL}$+GW, $\rm FRB_{\rm UL}$+$\rm FRB1_{\rm L}$, and $\rm FRB_{\rm UL}$+$\rm FRB2_{\rm L}$ data. Here, CBS stands for CMB+BAO+SNe.}\label{Fig:compare_FRBUL+FRBL}
\end{figure*}

As the main focus of this paper, we report the constraint results from the full FRB dataset of both unlensed and lensed events, i.e., $\rm FRB_{\rm UL}$+$\rm FRB1_{\rm L}$ and $\rm FRB_{\rm UL}$+$\rm FRB2_{\rm L}$ in Table~\ref{tab:5}, which represent multiple probes from late-universe observations alone.

Here we emphasize two advantages of jointly analyzing the unlensed and lensed FRB data: 
(i) It serves as pure late-universe probes, thereby avoiding the impact of the tension between the early and late universe.
(ii) It holds promise for precisely constraining key cosmological parameters, where unlensed and lensed FRBs can give tight constraints on dark energy and the Hubble constant, respectively, as shown in sect.~\ref{ULL alone}. In upcoming FRB observations, e.g., of the CASM, a large detected FRB sample may harbor lensed bursts that have yet to be identified, so the joint method leads to an in-depth analysis that explores the cosmological value of the future FRB sample. 

It is evident that the FRB sample with lensed events can simultaneously provide precise measurements of $H_0$ and the EoS of dark energy. 
Remarkably, the addition of the unlensed events to the lensed events can significantly improve the constraint on $H_0$.
For example, in the $w$CDM model, the combinations of $\rm FRB_{\rm UL}$+$\rm FRB1_{\rm L}$ and $\rm FRB_{\rm UL}$+$\rm FRB2_{\rm L}$ achieve precision of $\varepsilon(H_{0})=1.0\%$ and $\varepsilon(H_{0})=0.4\%$, respectively, which are improved by about $47\%$ and $42\%$ compared with using the $\rm FRB1_{\rm L}$ and $\rm FRB2_{\rm L}$ data alone, respectively. 
However, the joint analysis does not effectively improve the constraints on dark-energy EoS parameters compared to using the unlensed FRB data alone. 
Quantitatively, in the $w_{0}w_{a}$CDM model, the constraints on $w_a$ from $\rm FRB_{\rm UL}$+$\rm FRB1_{\rm L}$ and $\rm FRB_{\rm UL}$+$\rm FRB2_{\rm L}$ are only about $7\%$ and $17\%$ better, respectively, than those from $\rm FRB_{\rm UL}$ alone.
Note that in the $w$CDM model, the improvements are even less significant. This is because the large number of unlensed FRBs already provide strong constraints on dark energy, weakening the contributions of additional probes.
Figs.~\ref{Fig:FRBUL+FRBL}(a) and~\ref{Fig:FRBUL+FRBL}(b) show the posterior contours in the $w_{a}$--$H_0$ and $H_0$--$\Omega_{\rm b}h^2$ planes, respectively, by using different FRB datasets. The different orientations of the contours constrained by $\rm FRB_{\rm UL}$ and lensed FRBs allow for mutually breaking parameter degeneracies. Obviously, the effect is more significant for the parameter $H_0$ than for $w_a$, leading to very different improvements in the joint analysis.

To explore the cosmological potential of combining lensed and unlensed FRBs, we compare the results with those from combining FRB datasets with low-redshift cosmological probes like BAO, SNe, and GW.
The BAO+SNe combination represents current optical probes, while GW serves as a promising non-optical probe. Future GW bright standard sirens can precisely constrain the Hubble constant, but effective constraints on dark-energy EoS parameters require supports from additional external probes. Ref.~\cite{Zhang:2023gye} have found that the synergy between future GW and FRB observations can achieve sub-percent precision on $H_0$. Consequently, we have included the constraints from $\rm FRB_{\rm UL}$+BAO+SNe and $\rm FRB_{\rm UL}+GW$ in Table~\ref{tab:5} for comparison.\footnote{We only selected the unlensed FRB data to represent combinations with non-FRB data, as the lensed FRB data provides less precise results.}
We also plot the two-dimensional constraint contours of CBS, $\rm FRB_{\rm UL}$+BAO+SNe, $\rm FRB_{\rm UL}$+GW for the $w_{0}w_{a}$CDM model in Fig.~\ref{Fig:compare_FRBUL+FRBL}, comparing them with contours from $\rm FRB_{\rm UL}$+FRB$1_{\rm L}$ and $\rm FRB_{\rm UL}$+FRB$2_{\rm L}$.
We first report the constraints of dark energy, and then those of the Hubble constant.

For constraining dark energy, any of the above combinations cannot significantly improve the constraints.
From Fig.~\ref{Fig:compare_FRBUL+FRBL}(a) showing the $w_0$--$w_a$ plane, we can see that the contours from $\rm FRB_{\rm UL}$+GW, $\rm FRB_{\rm UL}$+$\rm FRB1_{\rm L}$, $\rm FRB_{\rm UL}$+$\rm FRB2_{\rm L}$, $\rm FRB_{\rm UL}$+BAO+SNe, and CBS seem increasingly tighter.
When combining unlensed and lensed FRBs, the constraints provided by $\rm FRB_{\rm UL}$+$\rm FRB1_{\rm L}$, with $\sigma(w_{0})=0.096$ and $\sigma(w_{a})=0.79$, are very similar to those from the $\rm FRB_{\rm UL}$+GW, with $\sigma(w_{0})=0.095$ and $\sigma(w_{a})=0.80$.
Furthermore, $\rm FRB_{\rm UL}$+$\rm FRB2_{\rm L}$ offers slightly worse constraints than $\rm FRB_{\rm UL}$+BAO+SNe, with $\sigma(w_{0})=0.072$ and $\sigma(w_{a})=0.70$ for the former, versus $\sigma(w_{0})=0.068$ and $\sigma(w_{a})=0.61$ for the latter.
It is worth noting that the constraint on $w_0$ from $\rm FRB_{\rm UL}$+$\rm FRB2_{\rm L}$ is $14\%$ more precise than that from CBS.
We conclude that 
the inclusion of lensed FRBs in the unlensed sample could still deliver improvements of dark energy constraints, which is similar to including other late-universe probes: adding $5$ lensed FRBs is comparable to adding GW, and adding $40$ lensed FRBs is slightly weaker than adding BAO+SNe, but the $w_0$ constraint could still be better than that from CBS.

For constraining the Hubble constant, combining the lensed and unlensed FRBs can provide tighter constraints than other combinations.
In the $H_0$--$\Omega_{\rm b}h^2$ plane, Fig.~\ref{Fig:compare_FRBUL+FRBL}(b) shows increasingly tighter contours from $\rm FRB_{\rm UL}$+BAO+SNe, $\rm FRB_{\rm UL}$+$\rm FRB1_{\rm L}$, CBS, $\rm FRB_{\rm UL}$+GW, and $\rm FRB_{\rm UL}$+$\rm FRB2_{\rm L}$.
We give some examples in the $w_{0}w_{a}$CDM model.
We find that $\rm FRB_{\rm UL}$+$\rm FRB1_{\rm L}$ gives $H_0$ constraints similar to CBS, with precision of $\varepsilon(H_{0})=1.3\%$, compared to $\varepsilon(H_{0})=1.2\%$ from CBS.
Furthermore, $\rm FRB_{\rm UL}$+$\rm FRB2_{\rm L}$ delivers about a $36\%$ better constraint on $H_0$ compared to $\rm FRB_{\rm UL}$+GW.
Specifically, $\rm FRB_{\rm UL}$+$\rm FRB2_{\rm L}$ gives $\varepsilon(H_{0})=0.7\%$, meeting the standard of precision cosmology, while $\rm FRB_{\rm UL}$+GW offers about $\varepsilon(H_{0})=1.2\%$. 
These results show that, including $5$ lensed events in a sample of 100,000 FRBs can constrain $H_0$ with the precision similar to CBS, and including $40$ lensed FRBs could exceed that of including $1,000$ standard siren data from the future GW observation of ET.\footnote{The inclusion of the lensed events also significantly improve the constraint on $\Omega_{\rm b}h^2$ from Fig.~\ref{Fig:compare_FRBUL+FRBL}, which can help localized FRBs find ``missing'' baryons in the local universe.}

\begin{figure}[!bp]
\centering
\includegraphics[width=0.85\linewidth,angle=0]{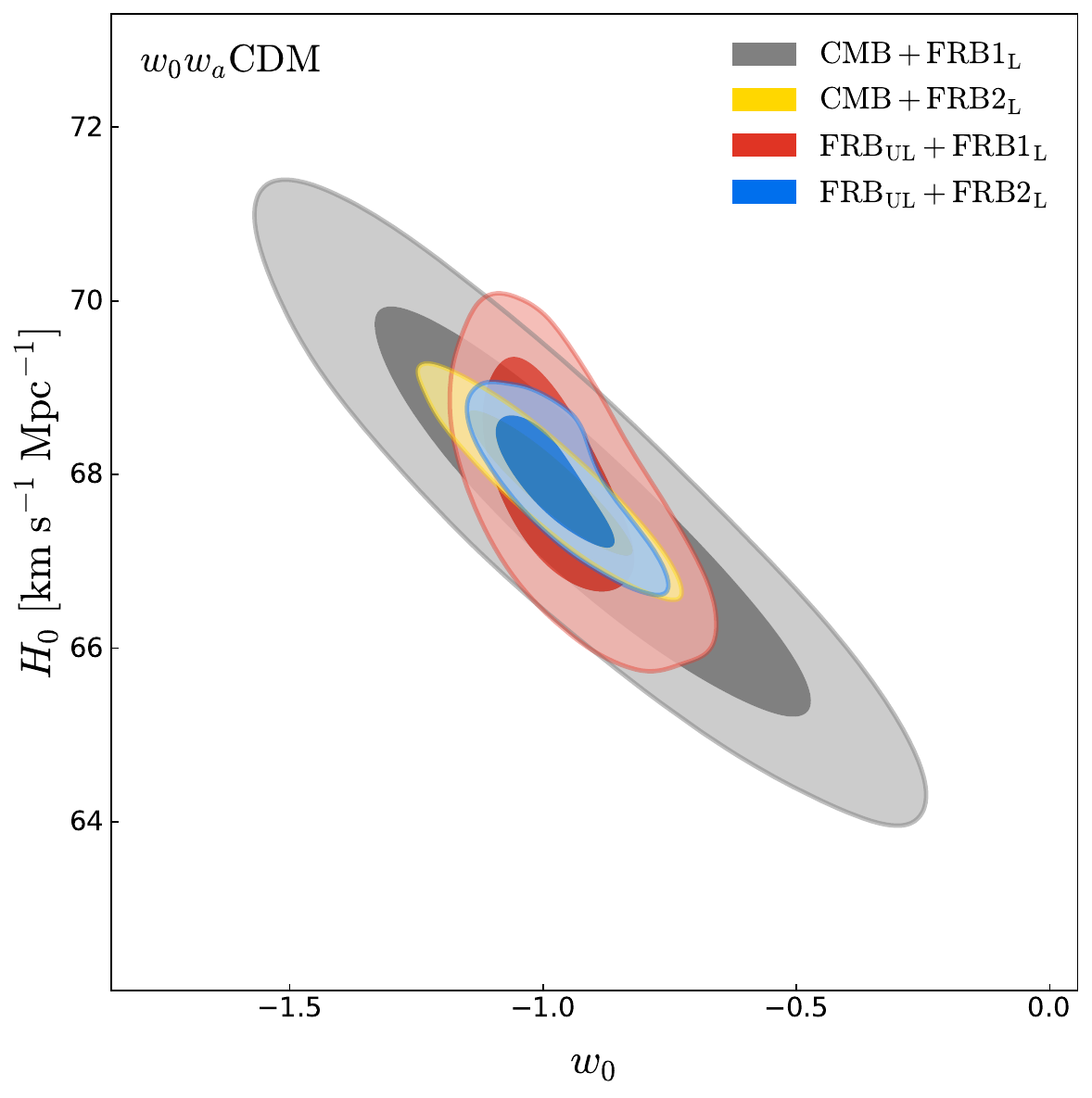}
\vspace{-0.3cm}   
\caption{Two-dimensional posterior distribution ($68.3\%$ and $95.4\%$ credible regions) in the $w_0$--$H_0$ plane for the $w$CDM model, by using the CMB+FRB1$_{\rm L}$, CMB+FRB2$_{\rm L}$, $\rm FRB_{\rm UL}$+$\rm FRB1_{\rm L}$, and $\rm FRB_{\rm UL}$+$\rm FRB2_{\rm L}$ data.} \label{Fig:compare_CMBL+FRBULL}
\end{figure}

Interestingly, we also compare the combined results of lensed FRBs with unlensed FRBs and CMB, and find that $\rm FRB_{\rm UL}$+$\rm FRB_{\rm L}$ could yield even more precise measurement of cosmological parameters than CMB+$\rm FRB_{\rm L}$ from Fig.~\ref{Fig:compare_CMBL+FRBULL}.
For example, the constraints on $H_0$, $w_0$, and $w_a$ from $\rm FRB_{\rm UL}$+$\rm FRB1_{\rm L}$ are $\sim 40\%$, $67\%$, and $12\%$ more precise, respectively, than those from CMB+$\rm FRB1_{\rm L}$ for the $w_{0}w_{a}$CDM models.
The constraints from combining both lensed and unlensed FRB data could be not only precise but also a late-universe result derived from the FRB observation alone. The dual advantages provide a potential cross-check against the ``Hubble tension''.

Overall, we conclude that the joint analysis of unlensed and lensed FRB datasets of CASM is highly valuable, which can significantly improve the precision in measuring the Hubble constant.
The effect of combining the lensed FRBs with the unlensed ones is better than combining the future GW observation (with the unlensed FRBs) , and combining the unlensed FRBs with the lensed ones could be more effective than combining the CMB observation (with the lensed FRBs).

\begin{table*}[!ht]
	\caption{Same as Table~\ref{tab:3} but using the the combined datasets from current and future cosmological observations, i.e., CBS+GW, CBS+$\rm FRB_{\rm UL}$+$\rm FRB1_{\rm L}$, and CBS+$\rm FRB_{\rm UL}$+$\rm FRB2_{\rm L}$ data. Here CBS stands for CMB+BAO+SNe. $H_0$ is in units of km s$^{-1}$ Mpc$^{-1}$.}
	\label{tab:6}
	\setlength{\tabcolsep}{7mm}
	\renewcommand{\arraystretch}{1.5}
	\begin{center}
		\begin{tabular}{ccccc}
			\hline \hline
			Model & Error & CBS+GW & CBS+$\rm FRB_{\rm UL}$+$\rm FRB1_{\rm L}$ & CBS+$\rm FRB_{\rm UL}$+$\rm FRB2_{\rm L}$ \\
			\hline
			\multirow{4}{*}{$w$CDM}
			& $\sigma(H_{0})$   & 0.51 & 0.30   & 0.21 \\
			& $\varepsilon(H_{0})$   & $0.8\%$ & $0.4\%$ & $0.3\%$ \\ 
			& $\sigma(w)$   & 0.028 & 0.017   & 0.017 \\
			& $\varepsilon(w)$   & $2.8\%$ & $1.7\%$ & $1.7\%$ \\ 
			\hline
			\addlinespace
			\multirow{5}{*}{$w_{0}w_{a}$CDM}
			& $\sigma(H_{0})$   & 0.62 & 0.38    & 0.29 \\ 
			& $\varepsilon(H_{0})$   & $0.9\%$ & $0.6\%$ & $0.4\%$ \\ 
			& $\sigma(w_{0})$   & 0.067 & 0.051   & 0.046 \\
			& $\varepsilon(w_{0})$   & $6.5\%$ & $5.1\%$ & $4.6\%$ \\ 
			& $\sigma(w_{a})$  & 0.22 & 0.16    & 0.15 \\
			\hline \hline
		\end{tabular}
	\end{center}
\end{table*}

\begin{figure*}[!ht]
	\begin{center}
		\includegraphics[width=0.95\linewidth,angle=0]{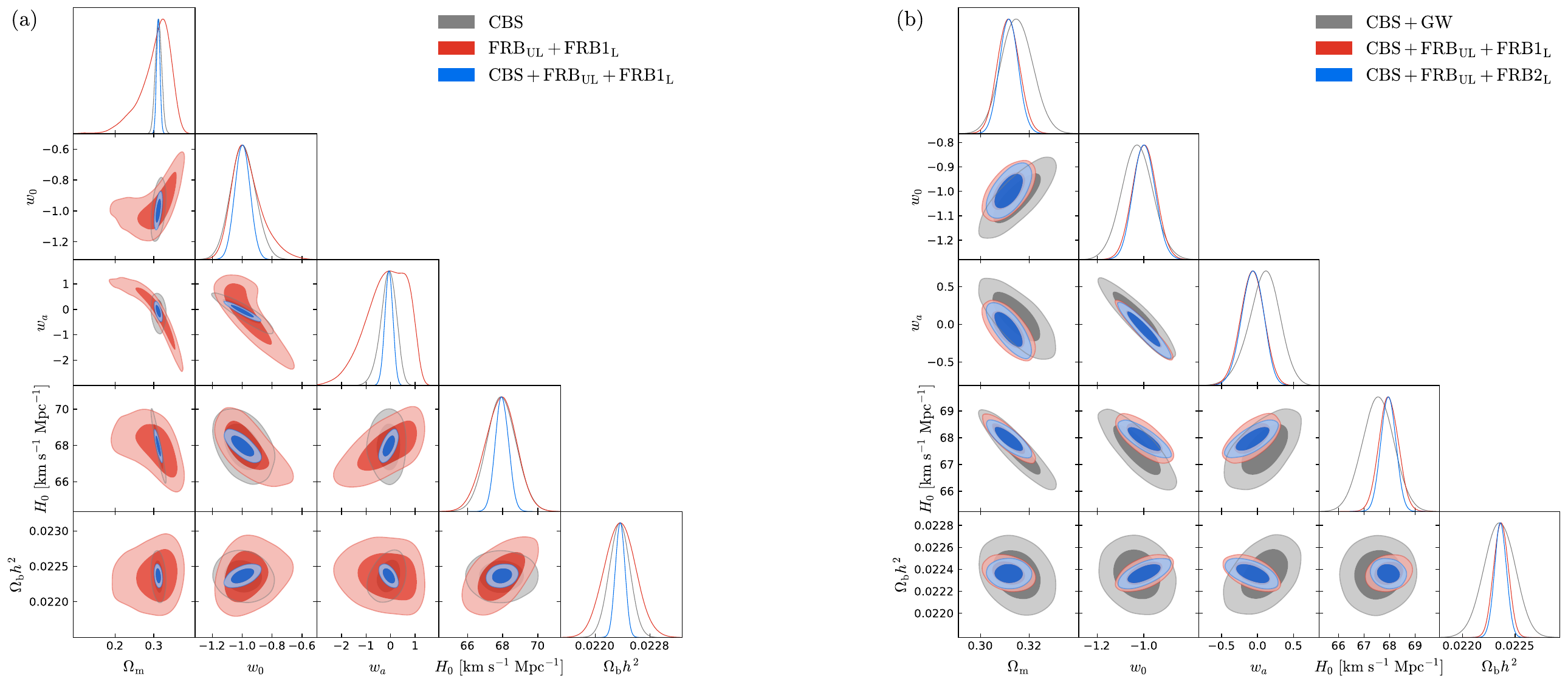}
	\end{center}
	\vspace{-0.3cm} 
	\caption{Three-dimensional posterior distribution ($68.3\%$ and $95.4\%$ credible regions)  for the $w_{0}w_{a}$CDM model, by using the CBS, $\rm FRB_{\rm UL}$+$\rm FRB1_{\rm L}$, and CBS+$\rm FRB_{\rm UL}$+$\rm FRB1_{\rm L}$ data (a) and the CBS+GW, $\rm FRB_{\rm UL}$+$\rm FRB1_{\rm L}$, and CBS+$\rm FRB_{\rm UL}$+$\rm FRB2_{\rm L}$ data (b).} \label{Fig:CBS+FRBUL+FRBL}
\end{figure*}

\subsection{Combination with the CBS data} \label{CBS+FRBULL}

Finally, we report the constraint results from combining the CBS data with the full FRB dataset, i.e., CBS+FRB$_{\rm UL}$+FRB1$_{\rm L}$ and CBS+FRB$_{\rm UL}$+FRB2$_{\rm L}$ in Table~\ref{tab:6}, which represent multiple probes from current and future cosmological observations.

We find that with the addition of $\rm FRB_{\rm UL}$+FRB$1_{\rm L}$ to CBS, the constraint precisions of $w_{0}$ and $H_0$ are improved from $9.9\%$ and $1.2\%$ to $5.1\%$ and $0.6\%$, respectively. 
For the parameter $w_a$, the absolute constraint error is significantly improved from $0.79$ to $0.16$, with a significant increase in precision of about $80\%$.
We plot three-dimensional posterior contours in Fig.~\ref{Fig:CBS+FRBUL+FRBL} for the $w_{0}w_{a}$CDM model.
As can be seen, the improvements are also evident for the parameters $\Omega_{\rm m}$ and $\Omega_{\rm b}h^2$.
Hence, the FRB observation of CASM will significantly improve current constraint precision of cosmological parameters.

To study the ability of the full FRB samples to improve cosmological parameter estimation, we compare the combinations CBS+$\rm FRB_{\rm UL}$+FRB$_{\rm L}$ with CBS+GW, which also represent multi-wavelength and multi-messenger observations, respectively. Meanwhile, we can thus investigate what precision the two perspectives of multiple probes may achieve in the future.
Constraint contours of CBS+GW, CBS+$\rm FRB_{\rm UL}$+FRB$1_{\rm L}$, and CBS+$\rm FRB_{\rm UL}$+FRB$2_{\rm L}$ for the $w_{0}w_{a}$CDM model are shown in Fig.~\ref{Fig:CBS+FRBUL+FRBL}.
We can clearly see that the contours from the joint CBS+$\rm FRB_{\rm UL}$+$\rm FRB_{\rm L}$ data are both tighter than those from CBS+GW. For example, the constraints from CBS+$\rm FRB_{\rm UL}$+FRB$2_{\rm L}$ include $\sigma(H_0)=0.29~{\rm km~s^{-1}~Mpc^{-1}}$, $\sigma(w_0)=0.046$, and $\sigma(w_a)=0.15$.
For the concerned parameters $H_0$, $w_{0}$, and $w_{a}$, the absolute errors from CBS+$\rm FRB_{\rm UL}$+$\rm FRB_{\rm L}$ are smaller than those from CBS+GW by about $39\%$--$53\%$, $24\%$--$31\%$, and $27\%$--$32\%$, respectively.\footnote{For CBS+GW, we also test the case of incorporating $10$ strongly lensed mock GW events with EM counterparts, following Ref.~\cite{Liao:2017ioi}, and find minimal improvement in parameter constraints (e.g., $H_0$), suggesting that the cosmological impact of strong gravitational lensing in GW bright siren observations is less significant compared to FRB observations.}

Overall, we conclude that when combining current cosmological probes, the FRB detections of the CASM may outperform the GW detections of the ET in constraining cosmological parameters. This reinforces the cosmological implication of a multi-wavelength observational strategy in optical and radio bands.

\section{Discussion}\label{sec4}

\begin{figure*}[!htbp]
	\centering
	\includegraphics[width=0.95\textwidth]{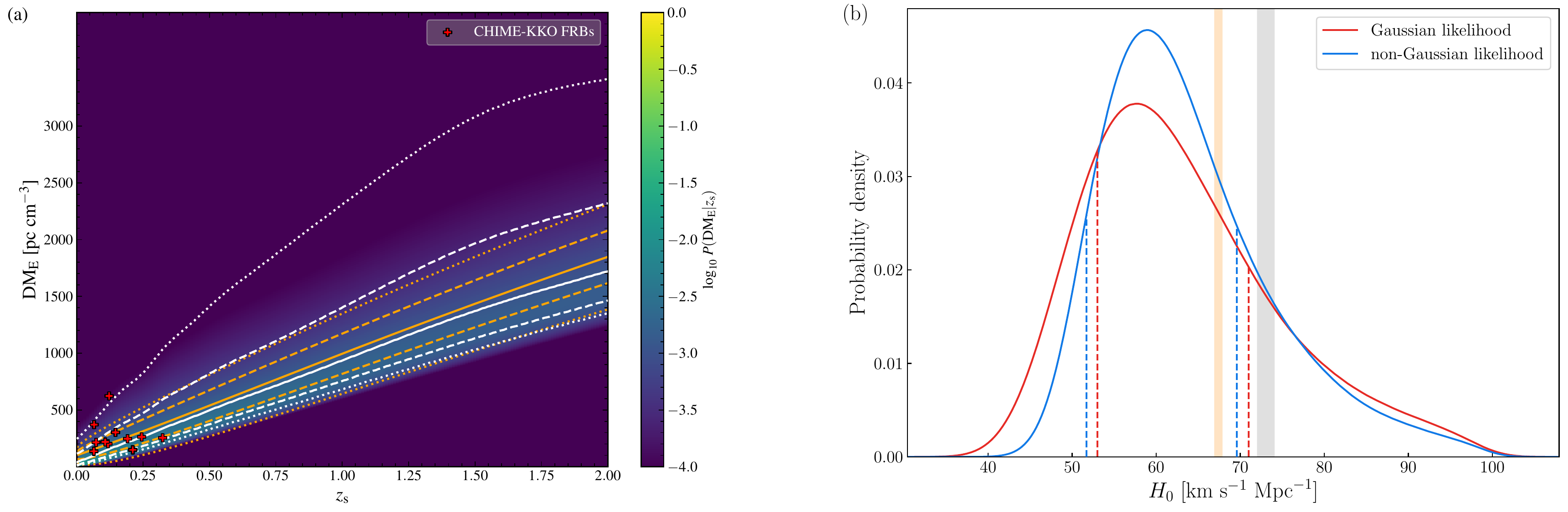}
	\centering
	\vspace{-0.3cm}  
	\caption{\label{fig:gauss_compare}
		Comparison between different likelihood frameworks for Gaussian and non-Gaussian frameworks. Panel (a) shows the distribution of $P({\rm DM_{\rm E}}|z_{\rm s})$. The Gaussian case assumes ${\rm DM_{\rm IGM}}$ with the mean value from Eq.~(\ref{aveDM}) and the standard error from Eq.~(\ref{eq:sigmadm}), and ${\rm DM_{\rm host}}$ plotted as $100~{\rm pc~cm^{-3}}$ (orange curve). 
		The non-Gaussian case assumes ${\rm DM_{\rm IGM}}$ and ${\rm DM_{\rm host}}$ following Eqs.~(\ref{pdf:igm}) and (\ref{pdf:host}), respectively, which is based on the IllustrisTNG simulation results \cite{Zhang:2020mgq,Zhang:2020xoc} (white curve).
		The color bar represents the posterior density for the non-Gaussian likelihood, while the solid, dashed, and dotted curves indicate the median, $68.3\%$, and $95.4\%$ credible regions for both cases. A dataset of 11 CHIME-KKO localized FRBs with secure host galaxy associations are shown as scatters \cite{Amiri:2025sbi} after subtracting the contributions of ${\rm DM_{\rm MW,ISM}}$ (using NE2001 model) and ${\rm DM_{\rm MW,halo}} = 55~{\rm pc~cm^{-3}}$. 
		Panel (b) shows the posteriors for $H_0$ within the $\Lambda$CDM model using the localized sample under the two likelihood frameworks. The Gaussian case and non-Gaussian case are displayed as red and blue solid lines, respectively, with their $68.3\%$ credible intervals marked by vertical dotted lines. The constraints from Planck \cite{2020Planck} and SH0ES \cite{Riess:2021jrx} are shown as orange and grey shaded regions, respectively.
	}
\end{figure*}
We now explore a more physics-based DM modeling framework
without assuming Gaussian distributions for $\rm DM_{IGM}$ and $\rm DM_{host}$.
This approach is particularly necessary given the limited number of localized FRB samples currently. 
We further compare this framework with the previous Gaussian model (i.e., non-Gaussian vs. Gaussian) and analyze the implications. Note that we only discuss the unlensed FRBs in this section.

In our original simulation, we assumed Gaussian distributions for both $\rm DM_{\rm IGM}$ and $\rm DM_{\rm host}$.
The realistic DM measurements exhibit non-Gaussian features due to the the IGM plasma inhomogeneity and host galaxy diversity.
To account for these effects, we revise our DM modeling framework following Ref.~\cite{Macquart:2020lln}.

For the IGM plasma inhomogeneity, the actual value of $\rm DM_{\rm IGM}$ would vary significantly around the mean value $\langle{\rm{DM}}_{{{\rm{IGM}}}}(z)\rangle$ due to the line-of-sight variance caused by intersecting foreground galaxy halos. We model $\rm DM_{\rm IGM}$ with a quasi-Gaussian distribution:
\begin{equation}\label{pdf:igm}
	P_{\rm{IGM}}(\Delta|A,C_0,\sigma_{\rm{DM}})=A \Delta^{-\beta} \exp \left[-\frac{\left(\Delta^{-\alpha}-C_0\right)^2}{2 \alpha^2 \sigma_{\rm{DM}}^2}\right],~\Delta>0,
\end{equation}
with $\Delta \equiv {\rm{DM}}_{\rm{IGM}}/\left\langle{\rm{DM}}_{{{\rm{IGM}}}}(z)\right\rangle$ and with $\alpha=3$ and $\beta=3$ assumed.
The distribution parameters $A$, $C_0$, and $\sigma_{\rm{DM}}$ are regarded as functions of redshift and interpolated with the IllustrisTNG best-fit results from Table 1 of Ref.~\cite{Zhang:2020xoc}.

For the host galaxy contribution, we adopt a quasi-Gaussian distribution for $\rm DM_{\rm IGM}$ in the source rest frame,
\begin{align}\label{pdf:host}
	P_{\rm{host}}
	&\left(\widehat{\rm{DM}}_{\rm{host}}|\mu_{\rm host},\sigma_{\rm host}\right)=
	\frac{1}{\sqrt{2\pi}\sigma_{\rm{host}}  \widehat{\rm{DM}}_{\rm{host}}} \nonumber\\
	&\times \exp \left[-\frac{\left(\ln \widehat{{\rm DM}}_{\rm{host}}-\mu_{\rm host} \right)^2}{2\sigma_{\rm{host}}^2}\right],
\end{align}
where the redshift-corrected host DM is defined as $\widehat{\rm{DM}}_{\rm{host}} = {\rm DM}_{\rm host}/(1+z)$.
The mean value and variance of the distribution are ${\rm e}^{\mu_{\rm host}}$ and ${\rm e}^{2\mu_{\rm host}+\sigma_{\rm{host}}^2}({\rm e}^{\sigma_{\rm{host}}^2}-1)$, respectively.
Ref.~\cite{Zhang:2020mgq} statistically estimated the median value of $\mathrm{DM_{\mathrm host}}$ for three types of FRBs from the IllustrisTNG simulations with a power-law redshift evolution: 
$\mathrm{DM_{\mathrm host}} = 34.72(1+z)^{1.08}$ pc cm$^{-3}$, 
$\mathrm{DM_{\mathrm host}} = 96.22(1+z)^{0.83}$ pc cm$^{-3}$, and
$\mathrm{DM_{\mathrm host}} = 32.97(1+z)^{0.84}$ pc cm$^{-3}$ for repeating FRBs in dwarf galaxies, repeating FRBs in spiral galaxies, and one-off FRBs, respectively, which are labeled as type 1, 2, and 3, respectively (see Appendix \ref{Appendix:tab} for more details).
Based on this classification, we calibrate the model parameters $\mu_{\mathrm{host}}$  and $\sigma_{\mathrm{host}}$ for each FRB using the fitting parameters from Table~3 of Ref.~\cite{Zhang:2020mgq}.

Then, the extragalactic DM within the non-Gaussian framework is obtained by convolving Eqs.~(\ref{pdf:igm}) and~(\ref{pdf:host}), leading to the joint likelihood:
\begin{align}\label{pdf:DMe}
	\mathcal{L}_{\rm NG}=
	&\prod_{i=1}^{N_{\mathrm{FRB}}}P_i\left(\mathrm{DM}_{\mathrm{E}, i} \mid z_{\mathrm{s}, i}\right) =\prod_{i=1}^{N_{\mathrm{FRB}}}\int_0^{\mathrm{DM}_{\mathrm{E}, i}} P_{\rm{host}}\left(\frac{\mathrm{DM}_{\rm{host}}}{1+z_{\mathrm{s}, i}}\right) \nonumber  \\
	&\times P_{\mathrm{IGM}}\left(\mathrm{DM}_{\mathrm{E}, i}-\frac{\mathrm{DM}_{\rm{host}}}{1+z_{\mathrm{s}, i}}\right) \mathrm{dDM}_{\text{host}}.
\end{align}
In Fig.~\ref{fig:gauss_compare}(a), we compare $P({\rm DM_{\rm E}}|z_{\rm s})$ within the Gaussian (orange) and non-Gaussian (white) frameworks.
While both frameworks are basically consistent within $1\sigma$ for $z \leq 2$ where CHIME/FRB can detect, the non-Gaussian framework spans a broader range at the $2\sigma$ level, indicating that it better accounts for low-redshift data with large $\mathrm{DM_{host}}$.

To assess the effect of these DM modeling frameworks on cosmological parameter estimation, we analyze a recent localized sample from the first operational CHIME/FRB k'niPatn k'l$\left._\mathrm{\smile}\right.$stk'masqt Outrigger (KKO) \cite{Amiri:2025sbi}. 
A ``gold sample'' of $19$ CHIME-KKO FRBs was identified with secure host galaxy associations at $>90\%$ confidence. 
We exclude sources where the modeled $\rm DM_{\rm MW,ISM}$ exceeds $30\%$ of the observed DM to reduce large uncertainties at low Galactic latitude.
Finally, we select 11 FRBs to
constrain $H_0$ within the $\Lambda$CDM framework (see Appendix \ref{Appendix:tab} for the catalog).
For convenience, we take Gaussian priors of $\mathcal{N}(0.317, 0.007^2)$ and $\mathcal{N}(0.02233, 0.00036^2)$ for $\Omega_{\rm m}$ and  $\Omega_{\rm b}h^{2}$, respectively \cite{2020Planck}.

The posteriors are shown in Fig.~\ref{fig:gauss_compare}(b). 
The constraints are statistically consistent, which gives $H_0=63^{+8}_{-10}$ and $H_0=63.7^{+5.9}_{-12.0}$ for the Gaussian and non-Gaussian cases, respectively and are both with $\sim 14.0\%$ precision. 
Although our sample does not exhibit an obvious bias, we caution that the bias  could become significant with larger samples. This is due to strong parameter degeneracies between the DM model parameters (e.g., those characterizing the log-normal $P_{\rm host}$
and the quasi-normal $P_{\rm IGM}$) and cosmological parameters, as previously reported by Refs.~\cite{Kalita:2024xae,Chen:2024ngo}.
We leave further studies using larger and high-$z$ CHIME/FRB localized samples to future work.

While \(P_{\rm IGM}\) has been well established from large-scale structure simulations, modeling \(P_{\rm host}\) remains challenging. To mitigate uncertainties in \(\mathrm{DM_{host}}\), two survey strategies can be pursued. First, since the host contribution becomes less dominant when $z \gtrsim 1.5$, incorporating high-$z$ FRBs into cosmological analyses can reduce host-related uncertainties in the total extragalactic DM. These sources can be detected by sensitive instrument like FAST. 
Second, the scattering timescale \(\tau_{\rm scatter}\) correlates with local plasma turbulence and provides an independent constraint on \(\mathrm{DM_{host}}\) without relying on any assumed distribution (e.g., log-normal or normal) \cite{Yang:2024vqq,Mo:2025wru}. Although current FRB samples with accurate \(\tau_{\rm scatter}\) measurements are limited, observations indicate that FRBs with profound \(\tau_{\rm scatter}\) values tend to reside in extreme magneto-ionic environments \cite{Niu:2021bnl}, which can be observed by high-time resolution telescopes like Australia's SKA Pathfinder (ASKAP).

We conclude that while accounting for DM host diversity and IGM inhomogeneity via non-Gaussian modeling improves the accuracy and reliability of DM measurements, it may also bias the cosmological parameter estimation due to the strong parameter degeneracies between the DM models and cosmology.
Using high-redshift FRBs and sources with measurable scattering timescale  can effectively reduce the systematic uncertainties in \(\mathrm{DM_{host}}\) modeling, thus improving the robustness of cosmological studies using FRBs.

\section{Conclusion}\label{sec5}
In this paper, we forecast cosmological parameter estimation using localized FRBs from the future sensitive CASM survey.
With continuous wide-field monitoring and high sensitivity, the CASM is expected to detect and localize a large sample of FRBs, which potentially include strongly lensed events by massive galaxies.
The study of FRB cosmology will greatly benefit from precise TD measurement of lensed bursts and DM information of abundant unlensed ones, via the time-delay cosmography and the Macquart relation, respectively.
We have employed both methods to study the cosmological potential of the simulated FRB sample with the CASM.
By using MCMC techniques, we have mainly focused on the constraints on the Hubble constant and dark-energy EoS parameters within the fiducial flat dynamical dark energy framework (i.e., the flat $w$CDM and $w_{0}w_{a}$CDM models). 
Based on different datasets, we have the following main findings.

(i) Using only the FRB dataset of {either unlensed or lensed} events as an independent late-universe probe, 
we found that a large number of 100,000 FRBs can effectively constrain dark-energy EoS parameters, with constraint error of $\sigma(w)=0.052$ in the $w$CDM model, 
and using $5$ and $40$ lensed FRB data can measure the Hubble constant with relative errors of $\varepsilon(H_0) = 2.4\%$ and $\varepsilon(H_0) = 1.0\%$ in the $w_{0}w_{a}$CDM model, respectively.

(ii) {Combining CMB with lensed FRBs} as multiple probes from both early- and late-universe observations, 
we found that the lensed FRB data can greatly improve the constraints over either single probe, by well breaking the parameter degeneracy induced by the CMB data.  
Compared to using the CMB and lensed FRB data alone, the joint analyses improve $H_0$ constraints by about over $80\%$ and $40\%$ for both $w$CDM and $w_{0}w_{a}$CDM models, respectively.
The effect of only $5$ lensed FRBs could be similar to that of current BAO+SNe in breaking the CMB degeneracies, and $40$ lensed FRBs could even be comparable to 100,000 unlensed FRBs.

(iii) Using the full FRB dataset of both unlensed and lensed events as multiple late-universe probes enables an in-depth cosmological analysis of the localized FRB samples, avoiding the impact of the tension between the early and late universe. 
We found that the combination can simultaneously constrain the Hubble constant and dark energy, with high precision of $\varepsilon(H_0)=0.4\%$ and $\varepsilon(w)=4.5\%$ in the $w$CDM model.
The inclusion of the lensed events to the unlensed ones can significantly improve the $H_0$ constraint; for example, including $5$ lensed FRBs can achieve the precision similar to CBS, and including $40$ lensed FRBs could exceed that of including other emerging late-universe probe like $1,000$ GW standard sirens from future ET's observation.
Although the constraints on dark energy are not significantly improved, the joint analysis could still offer improvements comparable to those from combining the unlensed FRBs with BAO+SNe.
Interestingly, the combination $\rm FRB_{\rm UL}$+$\rm FRB_{\rm L}$ could yield more precise constraints than CMB+$\rm FRB_{\rm L}$. 
Overall, the joint analysis of unlensed and lensed FRB datasets of the CASM is highly valuable, which can significantly improve the precision in measuring $H_0$. The synergy could be more effective than combining either the unlensed FRBs with future GW standard sirens or the lensed FRBs with CMB.

(iv) {Combining CBS with the full FRB datasets} as multiple probes from current and future cosmological observations, 
we found that the joint CBS and $\rm FRB_{\rm UL}$+$\rm FRB2_{\rm L}$ data give $\sigma(H_0)=0.29~{\rm km~s^{-1}~Mpc^{-1}}$, $\sigma(w_0)=0.046$, and $\sigma(w_a)=0.15$ in the $w_{0}w_{a}$CDM model.
The constraint results from the joint CBS+$\rm FRB_{\rm UL}$+FRB$_{\rm L}$ data are about $50\%$--$80\%$ tighter than those from CBS, and about $30\%$--$50\%$ than those from CBS+GW. 
Therefore, the FRB observation of the CASM will significantly improve current constraint precision of cosmological parameters, which may outperform the GW detections of ET in constraining cosmological parameters. This reinforces the cosmological implications of a multi-wavelength observational strategy in optical and radio bands.

We discuss the implications of different DM modeling frameworks (Gaussian vs. non-Gaussian). No significant systematic biases are found using $11$ localized FRBs from the CHIME/FRB Outrigger KKO's gold sample. 
We give the caveat that the non-Gaussianity of DM models may introduce biases in cosmological parameter estimation (e.g., $H_0$). Incorporating high-$z$ FRBs and $\tau_{\rm scatter}$-inferred DMs can mitigate uncertainties in DM modeling and break degeneracies between DM and cosmological parameters.

This study aims to preliminarily explore the prospect of FRB cosmology, particularly considering the effect of strong gravitational lensing. 
It remains challenging to address the bias induced by systematic errors from the both methods and large-scale structure \cite{Reischke:2023blu,Takahashi:2024fkt}, as well as to observationally determine the FRB redshift from its host  \cite{Jahns-Schindler:2023wnz,Marnoch:2023mel}.
Nevertheless, future ambitious FRB observations are expected to resolve these issues and greatly contribute to deciphering the nature of dark energy, as well as resolving the Hubble tension if enough events with long-duration lensing are incorporated.

\begin{acknowledgments}
We are grateful to Wan-Peng Sun, Yichao Li, Tian-Nuo Li, Yun Chen, and Zheng-Xiang Li for helpful discussions. We thank Liam Connor for providing the \texttt{frb-grav-lensing} codebase \cite{Connor2022frbgravlensing}, which facilitated the lensing probability calculations. This work was supported by the National SKA Program of China (Grants Nos. 2022SKA0110200 and 2022SKA0110203), the National Natural Science Foundation of China (Grants Nos. 12473001, 11975072, 11835009, and 11875102), and the National 111 Project (Grant No. B16009).
\end{acknowledgments}

\appendix
\renewcommand\thefigure{\Alph{section}\arabic{figure}}
\renewcommand\thetable{\Alph{section}\arabic{table}}

\section*{Appendix}

\section{Lensing event rate estimation}\label{Appendix}
\begin{figure}[htbp]
	\centering
	\includegraphics[width=0.48\textwidth]{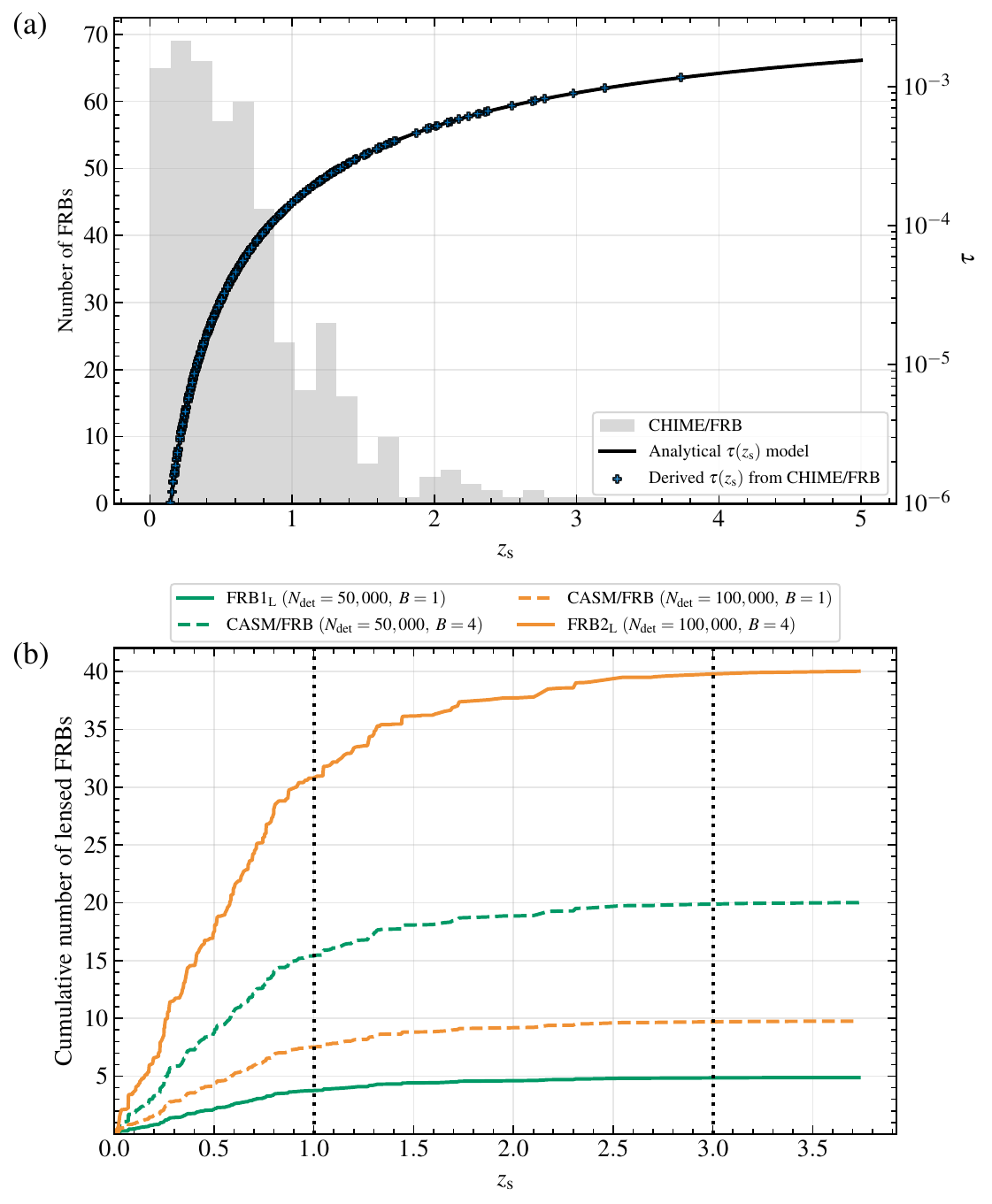}
	\centering 
	\caption{\label{fig:z_tau} {Calculation of the number of lensed FRBs detected by the CASM. Panel (a) shows the modeled CHIME/FRB redshift distribution as a grey histogram (left vertical axis), while the solid black line shows the lensing probability as a modeled function of source redshift and cross marks indicate  values derived from CHIME FRBs (right vertical axis).
			Adapted from \cite{Connor:2022bwl}.
			Panel (b) shows the cumulative number of lensed FRBs detected by CASM for sources with redshifts below $z_{\rm s}$, considering different assumptions on total detections and magnification biases. The vertical dotted lines show that the  lensed samples are accumulated to $\sim 80\%$ at $z_{\rm s} < 1$ and extended up to $z_{\rm s} \approx 3$.}}
\end{figure}

In this Appendix, we briefly summarize how the numbers of strongly lensed FRBs (i.e., 5 and 40) are predicted and the method for estimating the event rate of galaxy-galaxy lensing from Ref.~\cite{Connor:2022bwl}. The lensing probability calculations are calculated using the open-source code \cite{Connor2022frbgravlensing}.

To forecast the number of strongly lensed FRBs with the CASM, we resort to the lensing optical depth, which quantifies the probability that a source at redshift is gravitationally lensed:
\begin{equation}\label{eq:a1}
	\tau\left(z_{\rm s}\right)
	= \int_{0}^{z_{\rm s}}dz_{\rm l} \int_{0}^{\sigma_{v,{\rm max}}}d\sigma_vB(\gamma)\phi(\sigma_v, z_{\rm l}) \sigma (\sigma_{v},z_{\rm l},z_{\rm s})  \frac{d^2V}{d\Omega dz_{\rm l}}.
\end{equation}
Here $\sigma (\sigma_{v},z_{\rm l},z_{\rm s}) = \pi (\theta_{\rm E}D_{\rm l}^{\rm A})^2$ is the lensing cross-section. For a SIS lens, the Einstein radius is given by
\begin{equation}
	\theta_{\rm E} = 4\pi \frac{\sigma_v^2}{c^2} \frac{D_{\rm ls}^{\rm A}}{D_{\rm s}^{\rm A}},
\end{equation}
where $\sigma_{v}$ is the velocity dispersion of lensing galaxies. $\phi(\sigma_{v},z_{\rm l})$ is the velocity dispersion function (VDF) describing their number density. We adopt the analytical galaxy VDF from Eq.~(4) in Ref.~\cite{Yue:2021nwt}.
$\sigma_{v,{\rm max}} =1,000~\rm km~s^{-1}$.
$\frac{d^2 V}{d \Omega d z_{\rm l}}$ is the differential comoving volume at $z_{\rm l}$.
The parameter $B$ accounts for the magnification bias, which occurs when intrinsically faint sources below a survey's detection threshold are magnified by gravitational lensing and become observable. Given a power-law luminosity function, $\Phi(L) \propto L^{-\gamma}$, the bias can be parameterized as
\begin{equation}
	B(\gamma)=\frac{2^\gamma}{3-\gamma}.
\end{equation}
Since the slope $\gamma$ remains uncertain, we assume two cases based on current CHIME observation \cite{Zhang:2023pqs}: $\gamma = 1$ ($B = 1$) and $\gamma = 2$ ($B = 4$).

First, we derive the lensing optical depth for CHIME observations, since the CASM is hypothesized based on CHIME (which has CHIME/FRB's sensitivity but with $25$ times of the sky coverage). Using the first CHIME/FRB Catalog, we select 483 one-off sources at Galactic latitude above $5^{\circ}$. Their redshifts are estimated via a simplified Macquart relation (${\rm DM_{IGM}}(z) \approx 800z\rm~pc~cm^{-3}$ and $\rm DM_{host} =100~pc~cm^{-3}$). The modeled redshift distribution is shown in Fig.~\ref{fig:z_tau}(a).
Using Eq.~(\ref{eq:a1}), we derive $\tau(z_{\rm s})$ for each event, and the average optical depth is then computed as
\begin{equation}
	\bar{\tau} = \frac{1}{483} \sum_{i=1}^{483} \tau(z_{{\rm s},i}).
\end{equation}
Then, the expected number of lensed FRBs is estimated as
\begin{equation}
	N_{\rm L}= N_{\rm det}\bar{\tau},
\end{equation}
where $N_{\rm det}$ is the total number of detected FRBs from a given survey.

At last, we use CHIME FRBs' average lensing depth as a ``baseline'' to estimate $\tau(z_{\rm s})$ for CASM. The cumulative distributions of CASM's lensed FRBs under different assumptions of $N_{\rm det}$ and $B$ are shown in Fig.~\ref{fig:z_tau}(b).
The lensed samples accumulate rapidly although at low redshifts, reaching $\sim 80\%$ at $z_{\rm s} < 1$, suggesting good sample completeness even for low-redshift surveys.
This is because the lensing probability strongly depends on redshift, increasing rapidly as $z_{\rm s}^3$ for $z \leq 1.5$ in Fig.~\ref{fig:z_tau}(a). 
This accumulation extends up to $z \approx 3$. 
Current and future wide-area optical/IR surveys will serve as powerful tools for identifying galaxy-scale lenses, further improving sample completeness. 
The lower-bound case predicts $5$ lensed FRBs assuming a total detection of 50,000 FRBs and no magnification bias ($B = 1$), while the upper-bound case estimates $40$ lensed FRBs assuming 100,000 detected FRBs and strong magnification bias ($B = 4$).
Since we focus on what role the lensed bursts can play in large samples, we assume a total detection of 100,000 FRBs for convenience.
Thus, we consider two scenarios: a conservative scenario with $5$ lensed FRBs (labeled as $\rm FRB1_{\rm L}$) and an optimistic scenario with $40$ lensed FRBs (labeled as $\rm FRB2_{\rm L}$), both within a sample of 100,000 unlensed FRBs (labeled as $\rm FRB_{\rm UL}$).

\section{CHIME/FRB Localized Sample}\label{Appendix:tab}

\begin{table*}[!htbp]
	\caption{11 localized FRBs from CHIME/FRB Outrigger KKO's gold sample \cite{Amiri:2025sbi}.}
	\label{tab:chimekko}
	\setlength{\tabcolsep}{4.5mm}
	\renewcommand{\arraystretch}{1.5}
	\begin{center}
		\begin{tabular}{ccccccc}
			\hline \hline
			Name & Redshift & $\mathrm{DM}_{\mathrm{obs}}$ & R.A. (J2000) & Decl. (J2000) & $\mathrm{DM}_{\mathrm{MW,ISM}}$ & Type \\
			& & (pc cm$^{-3}$) & $(^\circ)$ & $(^\circ)$ & (pc cm$^{-3}$) & \\
			\hline
			FRB 20230203A & 0.1464& 420.1 & 151.66159 & 35.6941 & 67.299 & 3 \\
			FRB 20230222A & 0.1223& 706.1 & 106.96036 & 11.22452 & 33.31 & 3 \\
			FRB 20230311A & 0.1918 & 364.3 & 91.10966 & 55.94595 & 67.2352 & 3 \\
			FRB 20230703A & 0.1184 & 291.3 & 184.62445 & 48.72993 & 38.1566 & 3 \\
			FRB 20231017A & 0.245 & 344.2 & 346.75429 & 36.65268 & 31.1515 & 3 \\
			FRB 20231025B & 0.3238 & 368.7 & 270.78807 & 63.98908 & 65.5787 & 3 \\
			FRB 20231123A & 0.0729 & 302.1 & 82.62325 & 4.47554 & 37.1085 & 3 \\
			FRB 20231128A & 0.1079 & 331.6 & 199.5782 & 42.99271 & 64.7368 & 1 \\
			FRB 20231204A & 0.0644 & 221.0 & 207.99903 & 48.116 & 34.9372 & 2 \\
			FRB 20231206A & 0.0659 & 457.7 & 112.44284 & 56.25627 & 37.5114 & 3 \\
			FRB 20231230A & 0.0298 & 131.4 & 72.79761 & 2.39398 & 32.2305 & 3 \\
			\hline \hline
		\end{tabular}
	\end{center}
\end{table*}

In this Appendix, the physical properties of 11 localized FRBs used in sect.~\ref{sec4} are listed in Table~\ref{tab:chimekko}. The catalog includes redshift, observed DM, Right Ascension (R.A.), declination (Decl.), Galactic ISM DM derived from the NE2001 model, and host galaxy types. The host types are classified as type 1, 2, and 3, corresponding to repeating FRBs in dwarf galaxies, repeating FRBs in spiral galaxies, and one-off FRBs, respectively.

\bibliography{frbcasm}
\end{document}